\documentclass[letterpaper]{article}
\usepackage{natbib,alifeconf}

\usepackage{calc}
\usepackage[hyphens]{xurl}
\usepackage{hyperref}
\usepackage{tabularx}

\usepackage{authblk}
\usepackage{titlepic}
\usepackage{caption}
\usepackage{float}
\usepackage[T1]{fontenc} 

\graphicspath{ {images/} {images/fcd5/} }

\newcommand{\jargon}[1]{\textit{#1}}

\newcommand{\texsyn}[0]{TexSyn}
\newcommand{\lazypredator}[0]{LazyPredator}
\newcommand{\predatoreye}[0]{PredatorEye}

\newcommand{\runID}{\footnotesize}

\newcommand{\igfour}[1]{\includegraphics[width=0.24\linewidth]{#1}}
\newcommand{\igsix}[1]{\includegraphics[width=0.16\linewidth]{#1}}
\newcommand{\ignine}[1]{\includegraphics[width=0.104\linewidth]{#1}}

\newcommand{\stt}[1]{{\small \texttt{#1}}}

\begin{document}
\title{Coevolution of Camouflage}
\author{Craig Reynolds\authorcr
    unaffiliated researcher\authorcr 
    cwr@red3d.com}


\captionsetup{hypcap=false}

\titlepic{\igfour{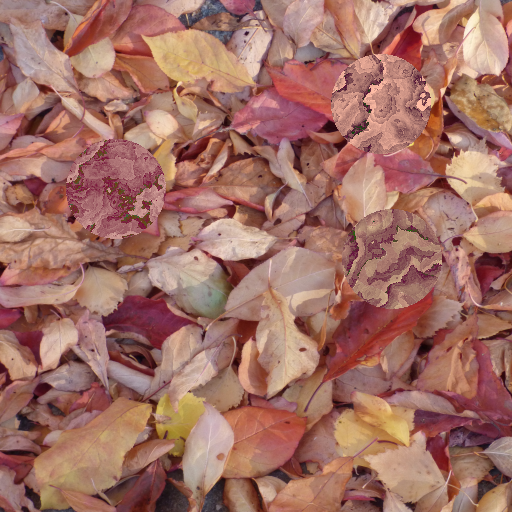} \hfill \igfour{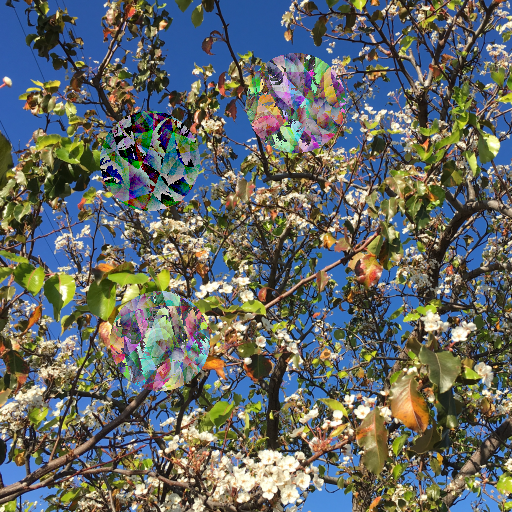} \hfill\igfour{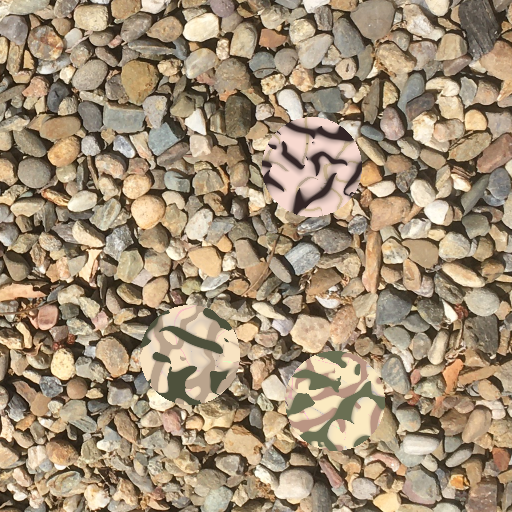}\hfill\igfour{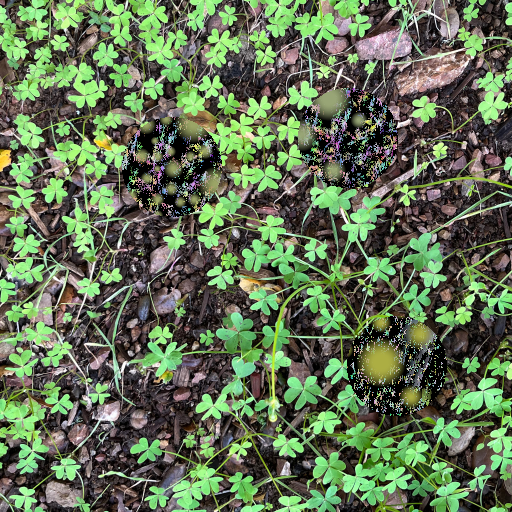} \captionof{figure}{Photographs of natural textures, each overlaid with three camouflaged \textit{prey}. The prey are randomly placed 2D disks, each with its own evolved camouflage texture. (Background photos of: plum leaf litter, tree and sky, gravel, oxalis sprouts. Zoom in for detail. Disk diameter is 20\% of image width.} 
\label{fig:teaser}}

\date{}

\maketitle



\begin{abstract}
    Camouflage in nature seems to arise from competition between predator and prey. To survive, predators must find prey, and prey must avoid being found. This work simulates an abstract model of that adversarial relationship. It looks at \textit{crypsis} through evolving prey camouflage patterns (as color textures) in competition with evolving predator vision. During their ``lifetime'' predators learn to better locate camouflaged prey. The environment for this 2D simulation is provided by a set of photographs, typically of natural scenes. This model is based on two evolving populations, one of prey and another of predators. Mutual conflict between these populations can produce both effective prey camouflage and predators skilled at ``breaking'' camouflage. The result is an open source \textit{artificial life} model to help study camouflage in nature, and the perceptual phenomenon of camouflage more generally.
\end{abstract}


\noindent{\small\textbf{Keywords:} camouflage, coevolution, nature, biology, predator, prey, vision, learning, texture synthesis, simulation, competition, adversarial. --- \textbf{Revised draft:} February 2025}


\section{Introduction}

\begin{figure*}[t]
    \igsix{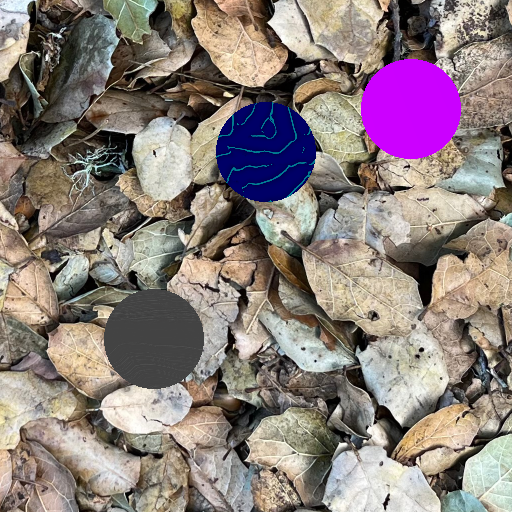}
    \hfill
    \igsix{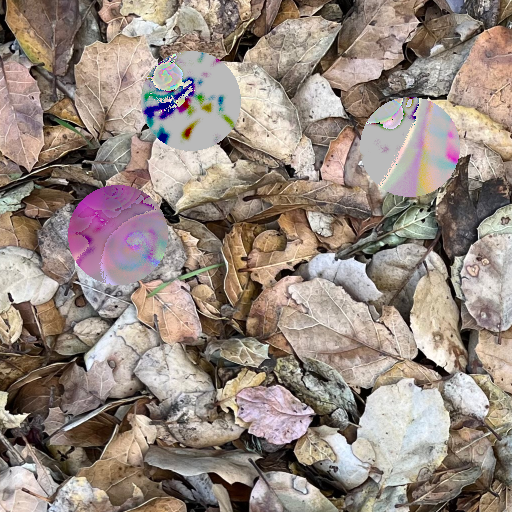}
    \hfill
    \igsix{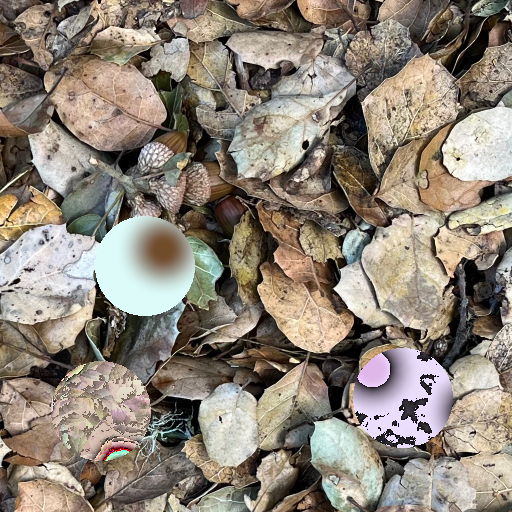}
    \hfill
    \igsix{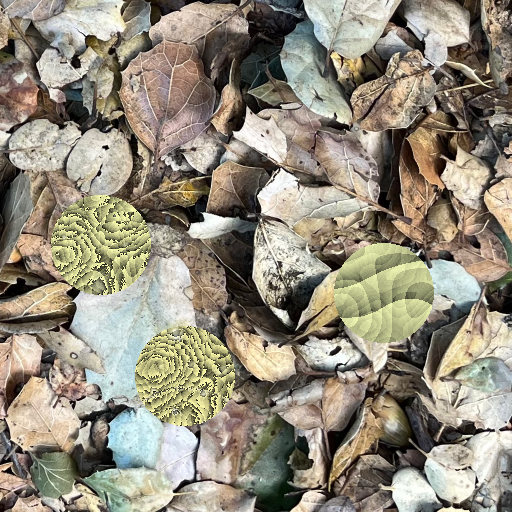}
    \hfill
    \igsix{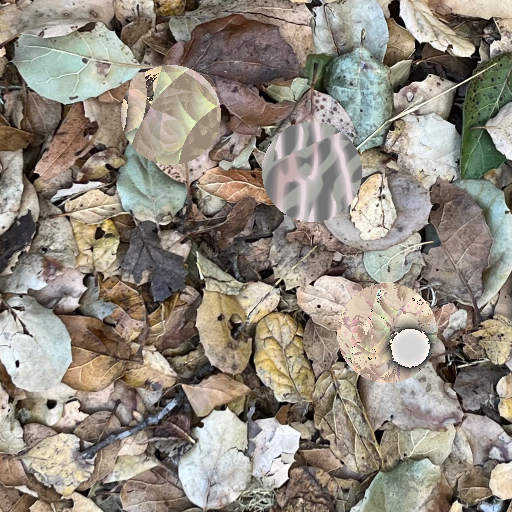}
    \hfill
    \igsix{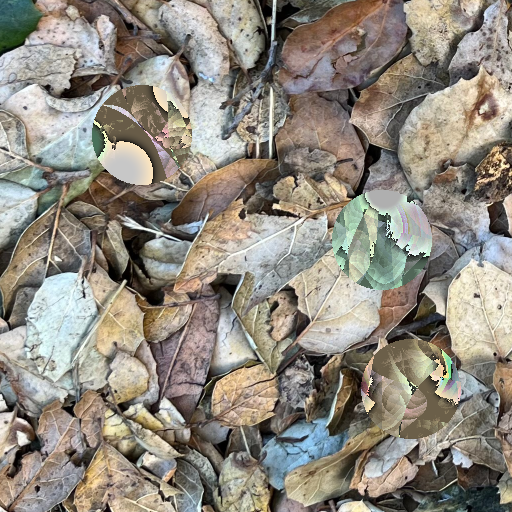}
    \caption{Prey camouflage evolving over simulation time to become more effective in a given environment.}
    \label{fig:time_sequence}
\end{figure*}

This work aims to create a simple abstract 2d simulation model of camouflage evolution in nature. These simulated camouflage patterns (Figure \ref{fig:teaser}) emerge from the interaction, the coevolution, of a population of simulated \jargon{prey} each with a candidate texture, and a population of simulated \jargon{predators} each with a learning visual detector. The simulation's main input is a set of photos of a background environment. Prey evolve to be \jargon{cryptic} (hard to find) against the background (Figure \ref{fig:time_sequence}). Evolving predators learn to hunt prey by locating their position within that 2D environment.
\par
Computational models of complex biological systems have several benefits. Constructing them, making them work as observed in nature, helps crystallize our thinking about natural phenomena. Computational models also allow experimentation \textit{in silico} to help understand these complex natural systems.
\par
This work follows the approach of an earlier simulation \citep{reynolds_iec_2011} where a population of camouflaged prey evolve in response to negative selection from a predator seeking conspicuous prey. In that earlier interactive game-like simulation, the predator was a human ``player.'' That simulation displayed a photographic background image overlaid with ten camouflaged prey. With five mouse clicks, the human predator selects the most conspicuous prey. Those selected prey, ``eaten'' by the predator, were removed from the population. Then they were replaced by \jargon{offspring} created with genetic \jargon{crossover} between surviving prey (the \jargon{parents}) followed by \jargon{mutation}. That simulation step was repeated about 2000 times.
\par


Here, evolution of prey camouflage closely follows that earlier work. While the human-in-the-loop is replaced with a population of predators, each based on a deep neural network. These CNN networks take an image as input and produce a \jargon{prediction}, an estimate, of where in the image the most conspicuous prey is located. See Figure \ref{fig:time_sequence} for an example of how this process unfolds over time.
\par
In abstract artificial life models, it is common to focus in detail on one aspect of a natural system. In the current model, that aspect is the coevolutionary dynamics between prey camouflage and predator vision. To make this feasible, other levels of organization are ignored, or assumed, or represented by a simple computational stand-in. So for example, the entire living organism that embodies these predators and prey, are simply assumed to exist, behaving as animals do, and are otherwise ignored. This model has a simple abstract representation of biological morphogenesis as programs (nested expressions) in \texsyn{}'s domain specific language for procedural texture synthesis \citep{reynolds_texsyn_2019}. \jargon{Genetic Programming} provides a simple model of evolution which acts on this ``genetic'' representation, creating new ``offspring'' textures through crossover and mutation. On the predator side, all of the animal's existence is ignored except for the key aspect of hunting behavior: looking at a scene and forming an opinion about where in the scene a prey is likely located. Learning from experience, these predators adapt to the appearance of an environment and the prey found there. Predators compete with each other on the basis of their ability to find prey, and so eat, and so survive. These details of simulated evolution, morphogenesis, vision, and genetic representation are all quite unlike the natural world. But they appear to be sufficiently similar in their effect to allow a plausible simulation of the natural system, producing analogous results, and so may provide insights about the natural system.
\par
To help ground this abstract simulation, consider a bird (predator), hunting for tasty but camouflaged beetles (prey), seen against the bark of a tree trunk (background image).
\par


\begin{figure*}[t]
    \includegraphics[width=\textwidth]{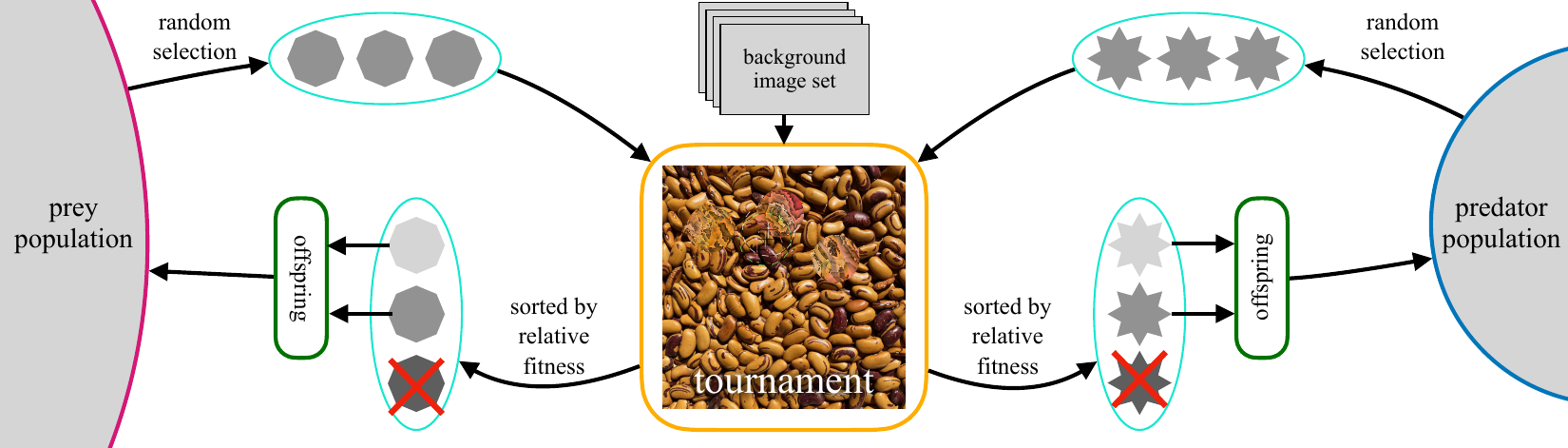}
    \caption{Overview of one step of the coevolutionary simulation of camouflage. Three prey are selected at random from their population of 400. Similarly for three predators from their population of 40. A random background image is selected from the given set, and a random crop of 512\textsuperscript{2} pixels is made. The three prey are rendered over the background at random non-overlapping locations. This composite \jargon{tournament image} is given to each predator which estimates a position (circled crosshairs in tournament image, see also Figure \ref{fig:predator_responses}) predicting the center point of the most conspicuous prey. The predators are scored by ``aim error'' --- the distance from their estimate to the \jargon{ground truth} center of the nearest prey. If the best predator's estimate is \textit{inside} a prey's disk, that prey is eaten and replaced by a new offspring of the other two prey. If all predators fail, all prey survive. If the worst scoring predator's estimate is \textit{outside} all prey, it may die of starvation, to be replaced by a new offspring predator.}
    \label{fig:simulation_overview}
\end{figure*}


\section{Related Work}
This work builds on \citet{reynolds_iec_2011} by replacing the human predator there with an evolving population of procedural predators, which \jargon{hunt} using a learning vision model. \citet{harrington_coevolution_2014} also used coevolution between prey and predators to create camouflage. Their predators detected prey with a multiscale convolution filter whose weights were evolved with a genetic algorithm.
\par
There was closely related work on learning surface textures to camouflage 3D objects within real 3D scenes. A technique for cubes is described in \citet{owens_camouflaging_2014}. Then \citet{guo_ganmouflage_2022} described an approach for arbitrarily shaped 3D objects. In both cases, the 3D scene is described by a set of photos from various viewpoints. The camouflage textures mapped onto these objects must trade off being inconspicuous from \textit{all} viewpoints in the scene. Recent work in field biology uses synthetic prey --- 3d printed then painted --- offered to wild predators to measure responses to specific aspects of camouflage \citep{kelley_role_2023}.
\par
Other computer graphics work related to camouflage include meticulously detailed reproduction of coloration patterns on real animals \citep{de_gomensoro_malheiros_leopard_2020}, generation of visual puzzles incorporating camouflaged images: \citep{chu_camo_image_2010}, \citep{Zhang_Yin_Nie_Zheng_2020}, a generative real time art exhibit based on mimicry \citep{wu_mimicry_2021}. CamoEvo \citep{hancock_camoevo_2022} is a toolbox for authoring online camouflage evolution games to understand evolution in real biological species. Such web-based simulations can be powered by very large numbers of human volunteers (''citizen scientists'') engaged in \jargon{human-based computation}.
\par
This adversarial coevolutionary simulation has clear similarities to \jargon{generative adversarial networks} (GANs) originally described in \citet{goodfellow_gan_2014}. Observer-driven optimization of camouflage patterns is an ideal application of GANs, as in CamoGAN \citep{talas_camogan_2020}. However the goal of the current work (as suggested in the Future Work section of \citet{reynolds_iec_2011}) is not \textbf{learning} camouflage, but to produce a simulation of a biological evolutionary system, suitable for ``what if'' experiments which cannot be performed in the natural world, e.g.: how does evolved camouflage change as the ratio of predators to prey changes? Some conjectures about natural camouflage might be better tested using A/B comparisons. For example, the relative value of specialist versus generalist camouflage (e.g. \citet{hughes_imperfect_2019}) or of background matching versus disruptive edge coloration (e.g. \citet{price_background_2019}). The simulation described here allows constructing carefully controlled experiments on these questions, without the difficulty of identifying comparable species in nature. Finally, this model demonstrates that, in the abstract, camouflage can arise in ``small'' populations over a ``short'' amount of time.
\par
The procedural texture synthesis used here to generate camouflage patterns has a long history. This work is perhaps most directly inspired by \citet{perlin_image_1985} where images are rendered from purely procedural representation of 3D textures. A recent example of this approach is found in \citet{Guerrero_MatFormer_2022}. Using texture synthesis under the control of a genetic algorithm goes back to \citet{sims_artificial_1991}, which in turn was inspired by the interactive \jargon{biomorph} evolution demo by \citet{dawkins_blind_1986}. FormSynth \citep{latham_form_1989} inspired Mutator \citep{todd_evolutionary_1994} and other tools. \citet{troscianko_quantifying_2017} attempt camouflage synthesis then evaluate effectiveness. Two other recent approaches to generating camouflage patterns are described in \citet{xiuxia_imitation_2023} and \citet{zhang_spatial_2013}. Indirectly related to camouflage, \textit{adversarial images} (which fool image classifiers into choosing the wrong category) have been created using genetic algorithms  \citep{bradley_generation_2023}. 
\par
\begin{figure*}[t]
    \includegraphics[width=\textwidth]{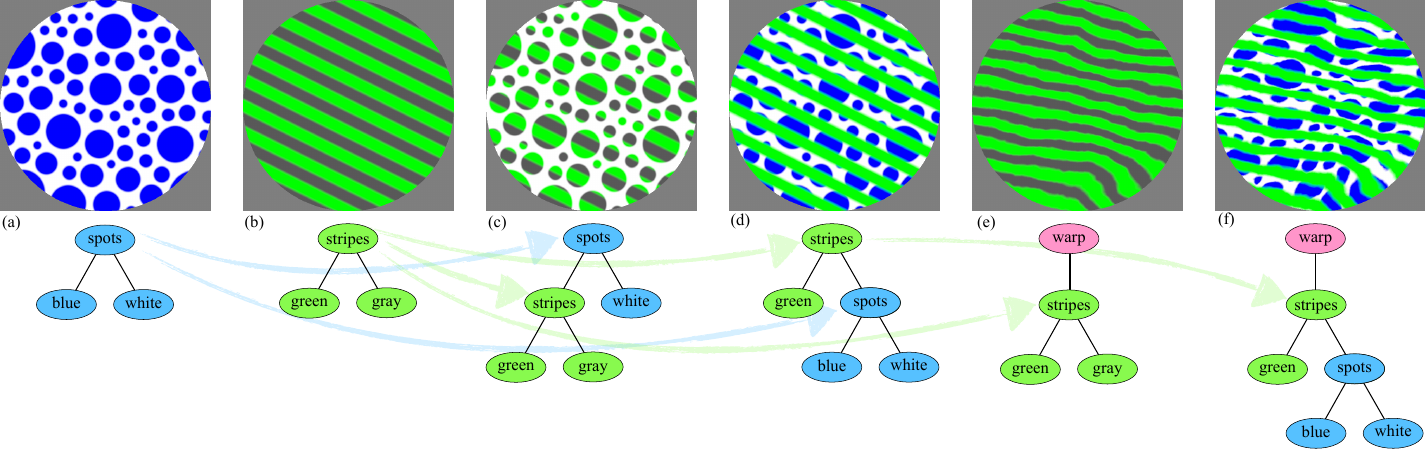}
    \caption{\texsyn{} expression trees and crossover between them, illustrated here with a simplified version of \texsyn{} with just three texture operators (\stt{spots}, \stt{stripes}, and \stt{warp}) plus four named solid color textures. Minimal operator trees are shown in (a) and (b). \textit{Crossover} between (a) and (b) is shown in (c) and (d). (c) is \stt{spots} where \stt{blue} is replaced with \stt{stripes}. (d) is \stt{stripes} where \stt{gray} is replaced with \stt{spots}. (e) and (f) show (b) and (d) under a \texttt{warp} operator. (See Section \nameref{sec:cpp_code} for the actual \texsyn{} c++ code used to create these examples.)}
    \label{fig:TexSyn_overview}
\end{figure*}

Evolution is represented in this model using \jargon{genetic programming} (GP), a population-based evolutionary optimization algorithm. It was first described by \citet{cramer_representation_1985} and popularized by \citet{koza_genetic_1992}. GP is a variation of \jargon{genetic algorithms} (GA) \citep{holland_genetic_1984}. GA traditionally use a fixed length bit string as its genetic representation, while GP uses an arbitrarily-sized tree-shaped representation. GP trees conveniently map onto nested expressions in a domain specific language. Texture synthesis in this work is based on nested expressions of texture operators from the \jargon{\texsyn{}} library, see Figure \ref{fig:TexSyn_overview}. A prey population of these textures is optimized for camouflage effectiveness by GP using the selection pressure from a population of predators which determine fitness. \texsyn{} is used with the \jargon{strongly typed} variant of Genetic Programming known as STGP \citep{montana_strongly_1995}, one of several grammar-based GP variants \citep{Mckay_2010}. GP has recently been used in the FunSearch work \citep{romera-paredes_mathematical_2023}. The GP implementation used here for camouflage evolution is called \jargon{\lazypredator{}} \citep{reynolds_lazypredator_2020}.
\par
The biological literature on camouflage and related topics is vast. A few starting points include: an authoritative modern survey \citep{cuthill_camouflage_2019}, a comprehensive early survey \citep{thayer_concealing-coloration_1909}, the influential book by \citet{cott_adaptive_1940}, pioneering work on mathematical models of biological patterns by \citet{turing_chemical_1952}, revisiting Turing's reaction-diffusion model with modern computation \citep{murray_how_1988}, and a contemporary perspective on how life evolves and learns \citep{valiant_probably_2013}. Also noteworthy are Endler's studies of camouflage, both experimental \citep{endler_natural_1980} and analytic \citep{endler_predators_1978}, \citep{endler_framework_2012}. Like Endler, \citet{brichard_natural_2023} look at interaction of camouflage and sexual signaling. A recent survey of how camouflage increases survival \citep{de_alcantara_viana_predator_2022} emphasizes how it need not be perfect. That a small change — increasing a predator's search time or reducing its attack rate — provides enough survival advantage to drive camouflage evolution.
\par


\subsection{Camouflaged Object Detection (COD)}
The last several years has seen a surge of computer vision research on \jargon{camouflaged object detection}, which simulates an aspect of predator behavior, specifically the \textit{breaking} of camouflage. (See this well-curated bibliography: \citet{visionxiang_cod}.) COD systems seek to \jargon{segment} camouflaged objects in images: identifying the pixels they cover. A recent example surveys this topic and presents a strong solution: \citet{Zhang2022}. Other research on COD include some based on boundaries \citep{chen_boundary-guided_2022}, \citep{sun_boundary-guided_2022}, a mixed-scale approach \citep{pang_zoom_2022}, one using transformer architecture \citep{yin_camoformer_2022}, and attempts to rank camouflaged objects by ``conspicuousness'' \citep{lv_cod_2022}, \citep{volonakis_camouflage_2018}.
\par
COD attempts \textit{a priori} camouflage ``breaking'' --- detecting the presence of well camouflaged objects --- without learning either the background or typical appearance of prey camouflage found in a given environment. That is, COD is a \jargon{generalist} predator, effectively using a form of \jargon{salience}. As summarized in \citet{Zhang2022}, COD is based on several labeled datasets (CHAMELEON, CAMO, COD10K, CAMO-FS \citep{nguyen_few-shot_2023}, ACOD2K dataset \citep{song_camouflaged_2023}) carefully annotated by hand at the pixel level. Newer work has looked at using generative techniques to create synthetic training data for COD \citep{zhang_camouflaged_2023}, and basing COD directly on a diffusion based approach: \citep{chen_diffusion_2023}, \citep{chen_camodiffusion_2023}, and with CLIP \citep{vu_leveraging_2023}. \citet{hu_shifting_2024} takes a novel approach, using virtual shadows as a form of co-supervision.
\par
In contrast, the goal of the simulation reported in this paper is to pit camouflage evolution against vision-based hunting. So determining the exact (pixel level) shape of the prey is not a requirement. This simulation ignores segmentation, abstracting prey as a disk of constant size, so sufficiently characterized by its center position. A real world predator does not require an exact segmentation to aim its attack at a prey's center. This work needs to find the \textbf{most} conspicuous prey, not all prey. This work simulates predators learning to find prey despite evolving camouflage patterns. This model \jargon{adapts} to dynamic camouflage rather than approach COD as a static task of generalist detection. Significantly, this work requires no hand-labeled training datasets at all, since it uses a form of \jargon{self-supervision}.
\par


\section{Components of the Simulation}

This section provides overviews of the various components that interact to form this coevolutionary camouflage model based on the interaction of predator and prey.


\begin{figure}[t]
    \centering
    \includegraphics[width=0.95\columnwidth]{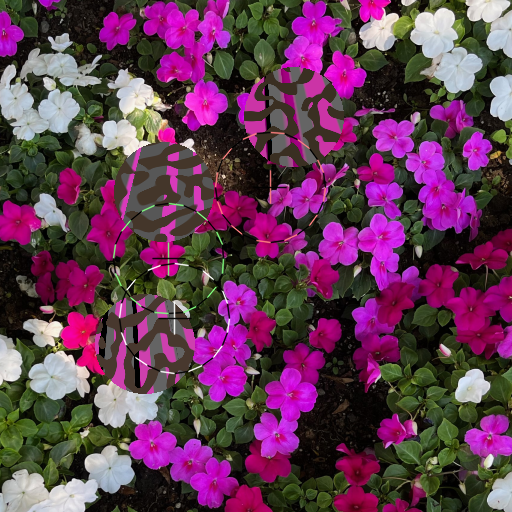}
    \caption{Tournament image after simulation step: three camouflaged prey on a random background crop. Three \jargon{crosshair} marks show the responses of three predators, ranked by minimum distance to a prey center. Details in Section \nameref{sec:additional_predator_responses}.}
    \label{fig:predator_responses}
\end{figure}

\begin{figure}[t!]
    \includegraphics[width=\columnwidth]{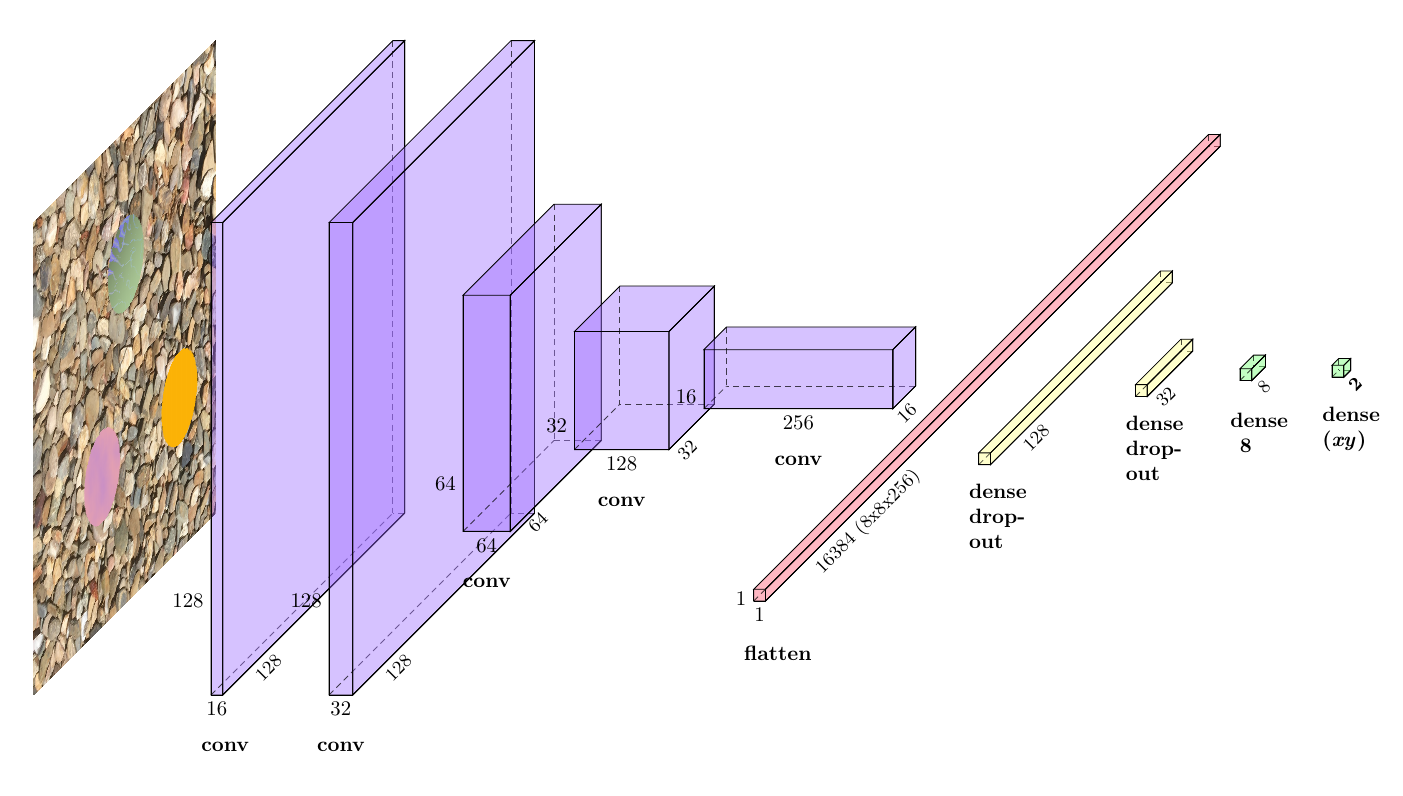}
    \caption{Architecture of a predator's neural net which maps a 128\textsuperscript{2} RGB image into an \textit{xy} location where it estimates the most conspicuous prey is centered. All convolution layers after the first use strides of (2,2) while doubling the number of learned filters. Then fully connected layers reduce the flattened features, by factors of 4, down to an output layer of 2 values. This model contains 3.2 million parameters (More details in Section \nameref{sec:pretrained_predator_details}.)}
    \label{fig:predator_cnn}
\end{figure}


\subsection{Coevolution, Populations, and Fitness}
This camouflage simulation is based on two adversarial \jargon{populations}: one of \jargon{predators} and one of \jargon{prey}. Individual prey compete for survival within their own population and similarly for predators. Predators must \jargon{hunt} successfully to ``eat'' and so survive. Prey survive if they are inconspicuous (cryptic) enough to avoid being found and eaten. A prey eaten by a predator, or a predator perished from hunger, is removed from its population. It is replaced by an \jargon{offspring} of parents from the surviving population.
\par
Predators define the fitness of prey: being easy to spot is bad, blending in is good. Similarly prey define the fitness of predators: being fooled by camouflage is bad, spotting cryptic prey is good. From this adversarial interaction, the two populations \jargon{coevolve}. If one side has some sort of flaw, the other side is motivated to exploit it. As a result both sides tend to improve over simulated time, see Figures \ref{fig:time_sequence}, \ref{fig:sqm_plot}, and \ref{fig:sqm_variance}.
\par
This notion of dynamic equilibria between coevolving species is often called the ``Red Queen hypothesis'' after \citet{van_valen_new_1973} who colorfully explains it with a quote from Lewis Carroll's \textit{Through the Looking Glass}: ``Now here, you see, it takes all the running you can do, to keep in the same place.''
\par
Initial random prey have coloration likely to contrast with the background. Initial predators have a \jargon{pre-trained} (''innate'') ability to find conspicuous (salient) objects which often allow them to hunt these initially un-camouflaged prey. They could be called \jargon{generalist} predators. As coevolution proceeds, prey become better camouflaged against the given background images. In response, predators learn to better hunt these prey on those backgrounds. They become \jargon{specialist} predators.
\par


\subsection{Tournaments, Competition, Relative Fitness}
\label{subsec:tournaments}
It is common in \jargon{evolutionary computation} to define \jargon{fitness} as a function that maps an \jargon{individual} (a member of the evolving population) into a number. That is, the function somehow evaluates the individual and assigns it a numeric score. Typically this fitness function is \jargon{idempotent}: its value depends only on static properties of the given individual. This can be seen as \jargon{absolute fitness}.
\par
In contrast, this simulation uses \jargon{relative fitness}, determined by \jargon{competition} between individuals, in \jargon{tournaments}. A tournament is a contest between multiple individuals \citep{angeline_competitive_1993}.
\par
A simple example of relative fitness is a foot race. The winner of the race is the first to cross the finish line. The order of finishing \jargon{sorts} the runners by speed. Races have been run this way since ancient times. Today it is simple to precisely measure each runner's absolute speed but that is not required to determine who won the race. Now consider two people playing chess. We do not know how to measure or predict a player's skill. But by pitting them against each other, having them play a game (or a series of games) the results provide a useful measurement of relative fitness. For a related study of fitnessless coevolution see \citet{jaskowski_fitnessless_2008}.
\par
Throughout this model, \textbf{tournaments involve three individuals}. (\citet{reynolds_iec_2011} used tournaments of size ten.) One simulation step (see Figure \ref{fig:simulation_overview}) consists of randomly selecting three prey individuals out of their population to compete in a tournament. Like a foot race, or chess game, a tournament serves to sort the individuals according to relative fitness. That relative fitness results from the behavior of adversarial predators. During the same simulation step, three predators are randomly selecting from their population. Each predator looks at the same input: an image with three camouflaged prey overlaid on a background image. Predators compete with each other by most accurately targeting prey. Prey compete with each other by hiding from (not being found by) the predators.
\par



\begin{figure*}[t]
    \begin{minipage}{\linewidth}
    \ignine{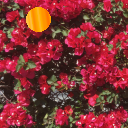}
    \hfill
    \ignine{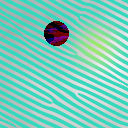}
    \hfill
    \ignine{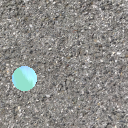}
    \hfill
    \ignine{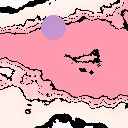}
    \hfill
    \ignine{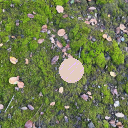}
    \hfill
    \ignine{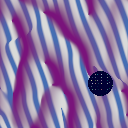}
    \hfill
    \ignine{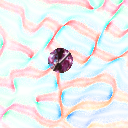}
    \hfill
    \ignine{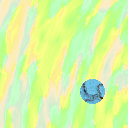}
    \hfill
    \ignine{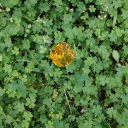}
    \end{minipage}
    \begin{minipage}{\linewidth}
    \vspace{0.1cm}
    \ignine{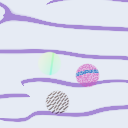}
    \hfill
    \ignine{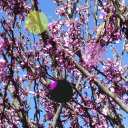}
    \hfill
    \ignine{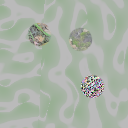}
    \hfill
    \ignine{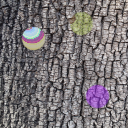}
    \hfill
    \ignine{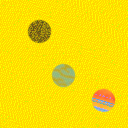}
    \hfill
    \ignine{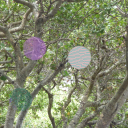}
    \hfill
    \ignine{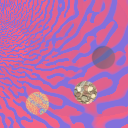}
    \hfill
    \ignine{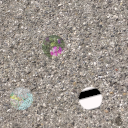}
    \hfill
    \ignine{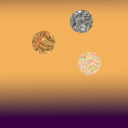}
    \end{minipage}
    \begin{minipage}{\linewidth}
    \vspace{0.1cm}
    \ignine{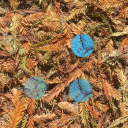}
    \hfill
    \ignine{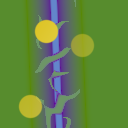}
    \hfill
    \ignine{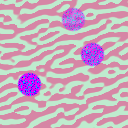}
    \hfill
    \ignine{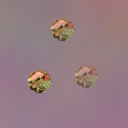}
    \hfill
    \ignine{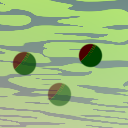}
    \hfill
    \ignine{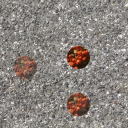}
    \hfill
    \ignine{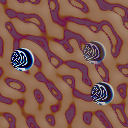}
    \hfill
    \ignine{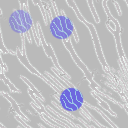}
    \hfill
    \ignine{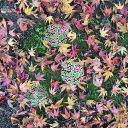}
    \end{minipage}
    \caption{Examples of the three types of labeled examples in the training dataset for ``find conspicuous disk'' task for  pre-trained CNN models FCD5 and FCD6 \citep{reynolds_FCD6_2022}. \textbf{Top row, type 1}: single random prey over random background. \textbf{Middle row, type 2}: three different prey. \textbf{Bottom row, type 3}: three copies of one prey. For types 2 and 3, one prey is unaltered, the other two are blended into background, by differing amounts, to make them more muted, and so perhaps less conspicuous. See Section \nameref{sec:pre_train_predator}.}
    \label{fig:fcd5_examples}
\end{figure*}


\subsection{Negative Selection, Drift, and Mixability}

A modern synthesis \citep{livnat_sex_2016} of evolution theory, game theory, and machine learning (the multiplicative weights update algorithm: ``no-regret learning'') suggests evolution may act to optimize \jargon{mixability}, a type of modularity in the genetic representation of phenotypes. \citet{chastain_multiplicative_2013} show it focuses on the ``...special case of weak selection in which all fitness values are assumed to be close to one another...hypothesizing that evolution proceeds for the most part not by substantial increases in fitness but by essentially random drift...'' This concept goes back to the ``Neutral Theory'' of \citet{kimura_evolutionary_1968}.
\par
\jargon{\lazypredator{},} the evolutionary model used here, operates in this ``neutral'' mode (as did \citet{reynolds_iec_2011}, \citet{harrington_coevolution_2014}, and others). Genetic algorithms are often designed to promote the population's ``best'' individuals. They allow these high fitness individuals to survive longer and reproduce more often. This \jargon{elitism} can skew the evolutionary search: too much \jargon{exploiting} without enough \jargon{exploring.}  In contrast, \lazypredator{} uses \textit{negative selection} to encourage \jargon{drift}. All high performing individuals are assumed to have similar fitness. It is the lowest fitness individuals that get culled by predation. (LazyPredator gets its name from this effect in nature: a lioness chases an antelope herd, causing its members to sort themselves by speed. The lion usually attacks the slowest individual at the back of the herd.) In each tournament (see Section \nameref{subsec:tournaments}) the goal is to sort the three individuals by relative fitness. In fact, the only requirement is to identify the \textbf{least fit} of the three individuals.
\par 
For example, a predator looks at a scene then predicts the location of the most conspicuous of three prey. If this location is within the disk-shaped ``body'' of a prey, it is the \textbf{least} fit of the prey tournament and gets ``eaten.'' The relative fitness ordering of the other two prey is not significant and is ignored. If all predators \jargon{fail} and predict positions outside all three prey, then the entire simulation step is abandoned, no prey is eaten, and the prey population is unchanged.
\par
In the predator population, a tournament is ranked by the \textbf{distance from a predator's prediction to the center of the nearest prey} --- essentially the predator's ``aiming error.'' The worst predator might then die from \jargon{starvation} based on its recent history of hunting success: has it eaten enough to survive? (Currently this threshold is 40\% hunting success over its last 20 attempts, see Table \ref{table:key_simulation_parameters}.) No specific limit on a predator's lifespans is imposed. Yet they tend to eventually die off, likely because the prey camouflage gets ``too good'' for them or they are out-competed by younger, better predators. This can be seen as a microscopic perspective on the ``Red Queen Hypothesis'' \citep{van_valen_new_1973}.


\begin{figure*}[t]
    \igfour{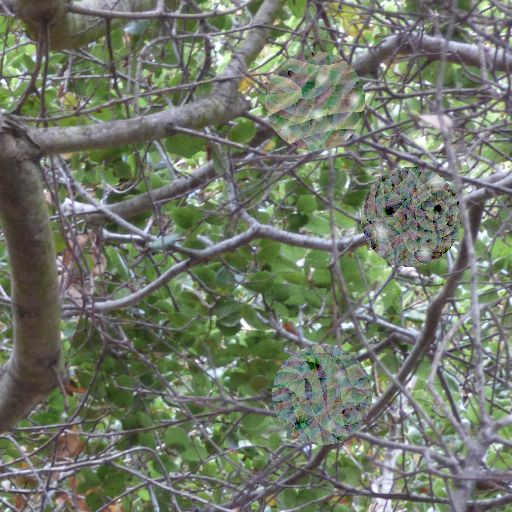}
    \hfill
    \igfour{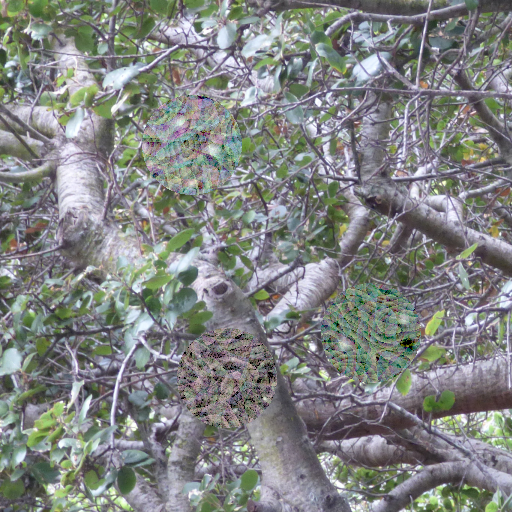}
    \hfill
    \igfour{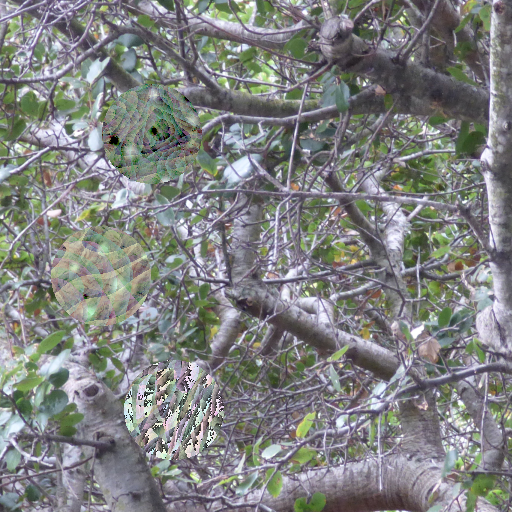}
    \hfill
    \igfour{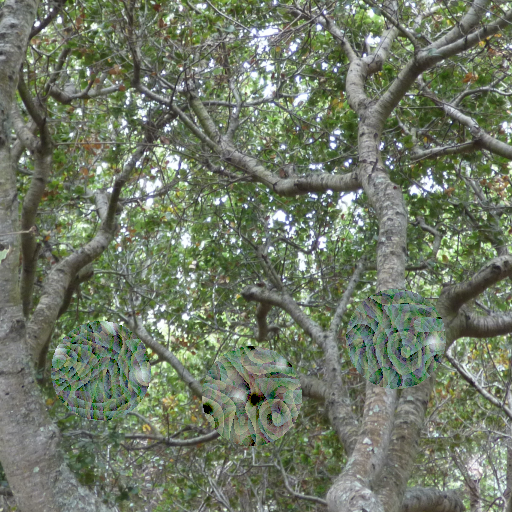}
    \caption{Four tournament images from run {\runID backyard\_oak\_20230113\_2254}.}
    \label{fig:backyard_oak_4x}
\end{figure*}

\begin{figure*}[t]
    \igfour{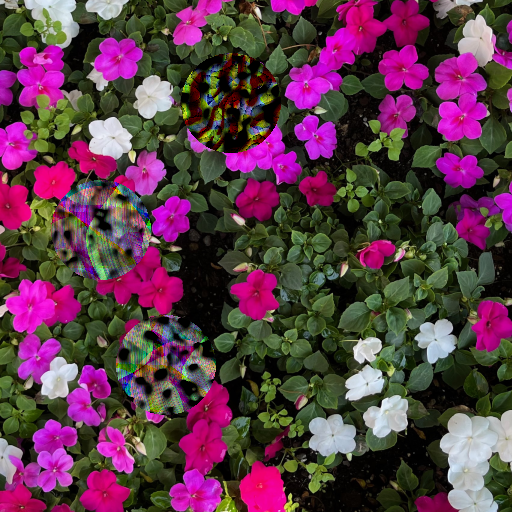}
    \hfill
    \igfour{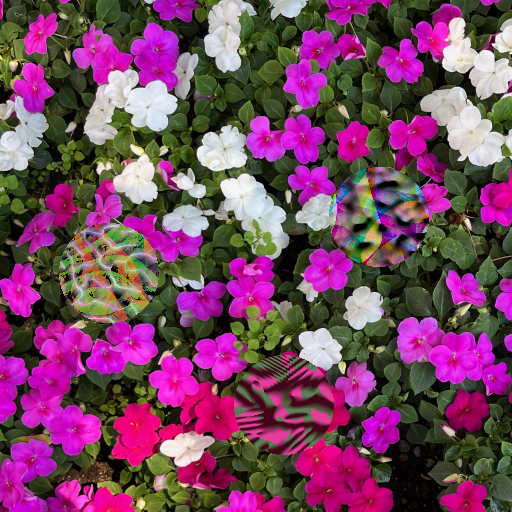}
    \hfill
    \igfour{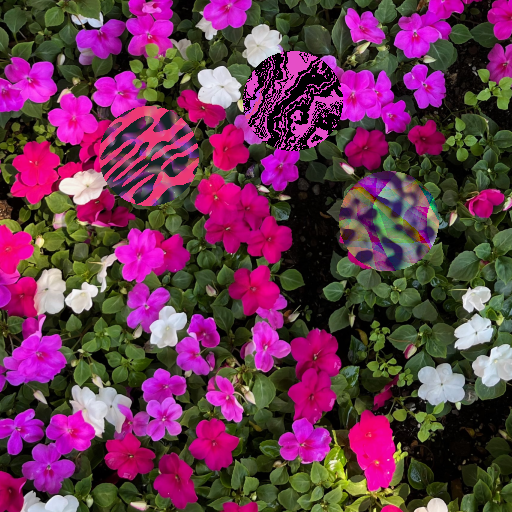}
    \hfill
    \igfour{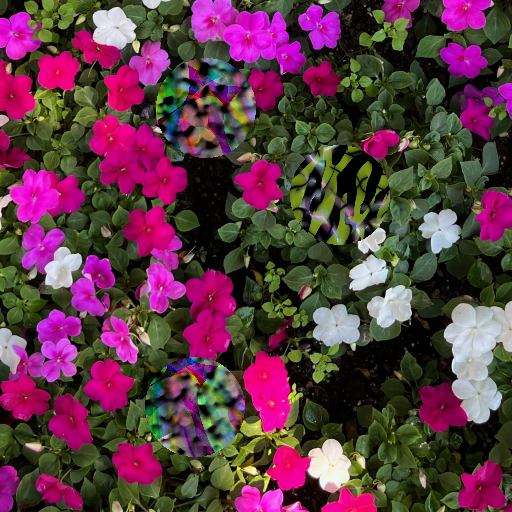}
    \caption{Four tournament images from run {\runID mbta\_flowers\_20230105\_1114}.}
    \label{fig:mbta_flowers_4x}
\end{figure*}


\subsection{Offspring, Crossover, and Mutation}
In the prey population, a tournament is used to find the lowest relative fitness (see Section \nameref{subsec:tournaments}). This corresponds to the worst camouflage, the \jargon{most conspicuous} of three prey in a tournament. If a predator successfully locates a prey, it is captured and ``eaten.'' The object representing that prey is removed from its population and replaced with a new individual (see Figure \ref{fig:simulation_overview}). This is the \jargon{population update} stage of a steady-state genetic programming system \citep{syswerda_study_1991}.
\par
The new prey, replacing the one eaten by the predator, is the \jargon{offspring} of two other prey. This is the motivation for using tournaments of size 3: the least fit prey dies, and is replaced by the offspring of the tournament's two surviving prey. This is why their relative fitness is irrelevant, they both become the \jargon{parents} given as input to the \jargon{crossover} operation.
\par
In GP (and \lazypredator{}) individuals are represented as tree structures. In this simulation, those trees are interpreted as \texsyn{} programs as described in Section \nameref{subsec:texture_synthesis}. The GP crossover operation is defined on two abstract parent trees. First they are copied to preserve the originals. One copy is chosen as recipient and one as donor. In each, a ``random subtree'' is selected. The recipient's random subtree is replaced by the donor's random subtree, by effectively splicing the pointers between tree nodes, see Figure \ref{fig:TexSyn_overview}.
\par
After crossover, a mutation operator further modifies the offspring. It traverses the tree, finding all the leaf nodes, which here correspond to numerical constants in the texture programs. Because \lazypredator{} uses \jargon{strongly typed genetic programming} \citep{montana_strongly_1995} each leaf belongs to a specific application-defined type (for example: a floating point number between 0 and 1). A method of the type's class can mutate (''jiggle'') the constant value of a leaf node (for example, add a small signed random offset, then clip back into the type's range). After crossover and mutation, the offspring is put into the population, replacing the dead prey.
\par
Predators are represented as convolutional neural nets, see section \nameref{subsec:predator_vision} below. For predator offspring, a surviving parent is selected from the tournament. All its parameters are directly copied, then mutated by adding signed noise to each model parameter. This is a population-based mutation-only \jargon{evolutionary strategy} as opposed to a \jargon{genetic algorithm} or \jargon{genetic programming}. Predators then learn (fine-tune) during their lifetime.
\par


\subsection{Texture Synthesis}
\label{subsec:texture_synthesis}
Camouflage textures are represented in this simulation as trees of procedural texture operators. These correspond directly to nested expressions in a typical programming language. \jargon{\texsyn{}} is a simple domain specific language for describing textures \citep{reynolds_texsyn_2019}.
\par
The details of \texsyn{} are not central to understanding this camouflage model. A quick overview is given here. \texsyn{} is a (c++) library, an API, with various \jargon{operators} each of which return a \jargon{Texture}. Most of them also take Textures as input parameters, along with simple values like colors, 2d vectors, and scalars. Nested expressions of \texsyn{} operators (''source code'') are compiled into trees of operator instances.
\par
These \texsyn{} textures are represented by operators, trees, and parameters --- but not pixel data. Instead the Texture class has a function to sample its color at any floating point \textit{xy} location. That color is computed on the fly, similar to GPU \jargon{fragment shaders} or the Pixel Stream Editor functions in \citet{perlin_image_1985}. Figure \ref{fig:TexSyn_overview} shows some simple examples and how these textures can be recombined with tree crossover.
\par
Note that the images in this paper are rendered at 512\textsuperscript{2} pixels and the prey disks have a diameter of 100 pixels. These are downsampled to 128\textsuperscript{2} for use by predator's vision CNNs. 
\par


\begin{figure*}[t]
    \igfour{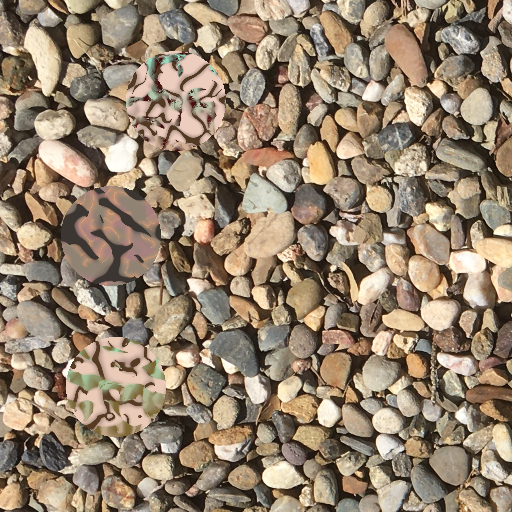}
    \hfill
    \igfour{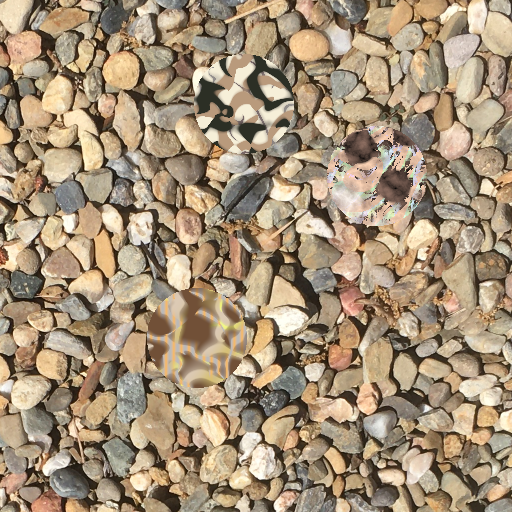}
    \hfill
    \igfour{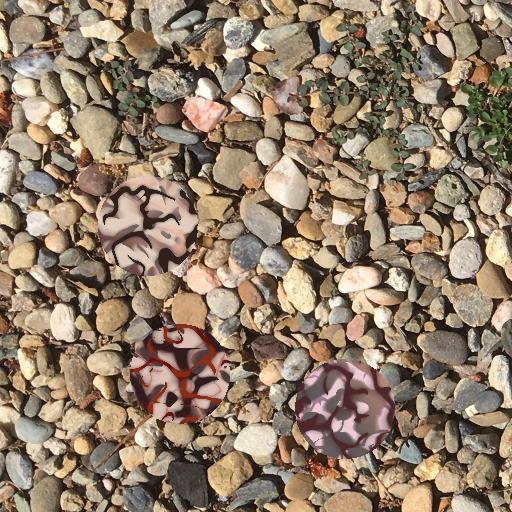}
    \hfill
    \igfour{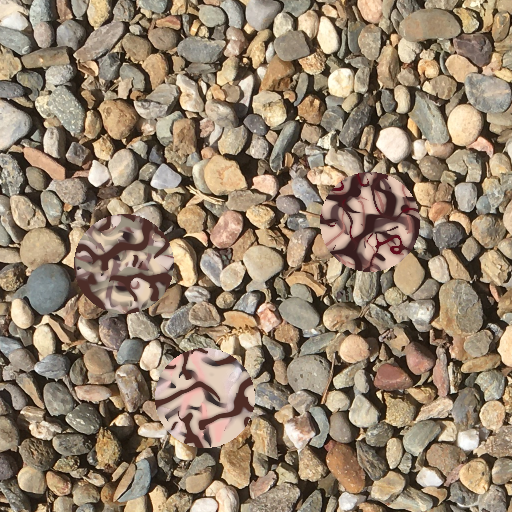}
    \caption{Four tournament images from run {\runID michaels\_gravel\_20221214\_1837}.}
    \label{fig:michaels_gravel_4x}
\end{figure*}

\begin{figure*}[t]
    \igfour{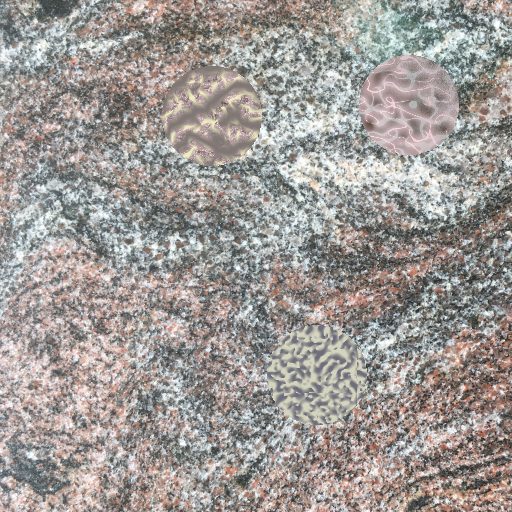}
    \hfill
    \igfour{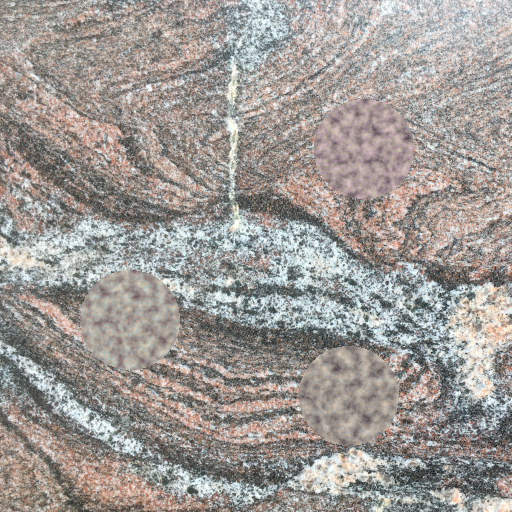}
    \hfill
    \igfour{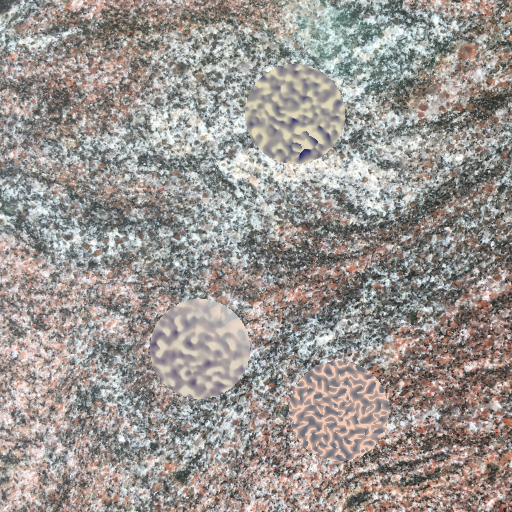}
    \hfill
    \igfour{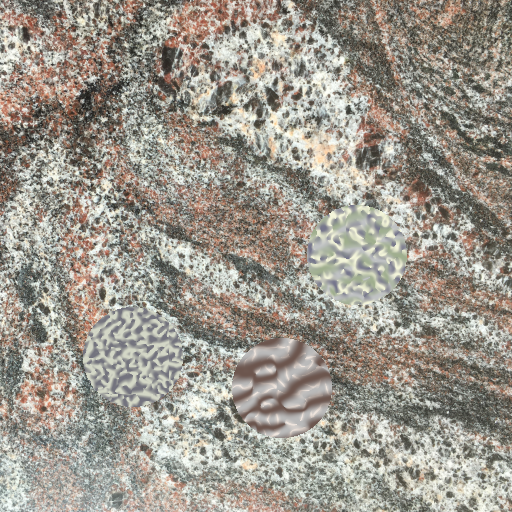}
    \caption{Four tournament images from run {\runID kitchen\_granite\_20230110\_1758}.}
    \label{fig:kitchen_granite_4x}
\end{figure*}


\subsection{Predator Vision}
\label{subsec:predator_vision}
It is the predator's job to look at a tournament image and ``hunt'' for prey. These images are built from a portion of a background photo and overlaid with three randomly placed camouflaged prey. The camouflage texture for each disk shaped prey is rendered on the \texsyn{} side. Because all images in this simulation are synthetic, they are labeled with the random \jargon{ground truth} position data for each prey. This allows predators to learn in a \jargon{self-supervised} manner.
\par

\subsubsection{Pre-Training Predator's Vision}
\label{sec:pre_train_predator}
The basis for a predator's visual system is \textbf{a \jargon{pre-trained} deep neural net model for a ``find conspicuous disk'' task}, see Figure \ref{fig:predator_cnn}. The goal of this task is to look at an arbitrary image and locate the centerpoint of the most conspicuous (salient) region, assumed to be a prey-sized disk. This pre-training is done once then reused for each subsequent camouflage evolution run. \texsyn{} was used to generate a dataset of 20,000 labeled training examples called \jargon{FCD6} \citep{reynolds_FCD6_2022}, see Figure \ref{fig:fcd5_examples}. Using augmentation by random variation, the effective training set size was 500,000. Each example was an RGB image, a 128$\times$128$\times$3 tensor, with an associated label: an \textit{xy} coordinate pair indicating a location in the image.
\par
Each training image starts with a \jargon{random texture} or a random crop of a photo (from a library of background images, see Section\nameref{subsec:background_sets}) over which are one or three prey disks with random texture. Although ``random texture'' is a slippery concept, the meaning here is the kind of prey texture used to initialize the prey population before an evolution run. The \lazypredator{} genetic programming engine has the ability to create random trees of a given size from a user-defined \jargon{function set} such as one for \texsyn{}. A random tree is interpreted as a random nested expression of \texsyn{} operators with randomly chosen leaf constants. As such these prey textures are quite varied, but have a fair amount of structure, they are not uncorrelated noise. See leftmost image in Figure \ref{fig:time_sequence}.
\par

Each image in the ``find conspicuous disk'' training dataset is generated in one of three styles chosen with equal probability. Style 1 has a single prey disk, the label is its center point. Style 2 has three different prey, style 3 has three copies of one prey disk. The latter two cases reduce visibility of two disks, by blending or dithering pixels of the prey disk into the background. The label corresponds to the unchanged disk, presumed to be more conspicuous.
\par

\subsubsection{Fine-Tuning Predator's Vision}
During simulations of camouflage evolution, each predator is initialized as a copy of the pre-trained ``find conspicuous disk'' model to which noise is added. (Zero mean noise of less than ±0.003 is added to each parameter of the deep neural net.) In each simulation step (see Figure \ref{fig:simulation_overview}) the three predators chosen to participate in the tournament predict a prey position. Then \textbf{each predator is \jargon{fine-tuned} based on a dataset of labeled images collected during this simulation run}. 
Each predator collects its own dataset consisting of tournament images in which it participated, essentially its memory of prey coloration in this environment. The fine-tuning dataset starts empty, then collects each tournament image until the dataset holds 500 images, then replaces one of the 500 chosen at random. Labels for a predator's dataset are the predictions it made in each tournament. See Figure \ref{fig:predator_responses}. The predator's CNN is fine-tuned (trained) on this set of (up to) 500 images. The samples are initially in temporal order, then become randomized as the predator survives more than 500 tournaments. This approach is analogous to the use of \jargon{batching} in \jargon{deep Q-learning}. See for example Algorithm 5.1 in \citet{casgrain_deep_2022}.
\par


\begin{figure*}[t]
    \igfour{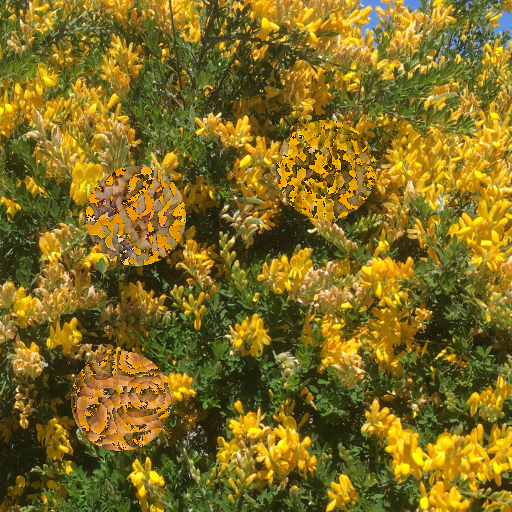}
    \hfill
    \igfour{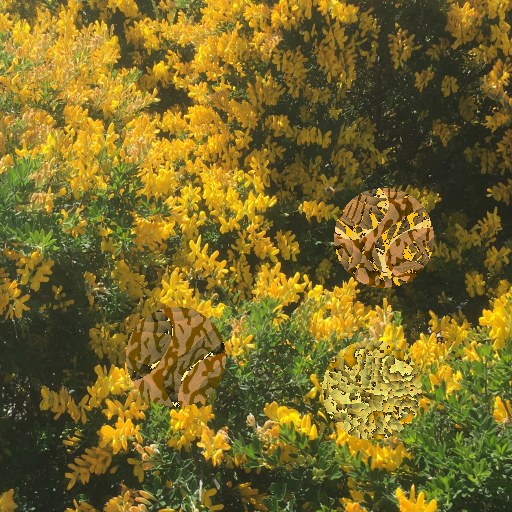}
    \hfill
    \igfour{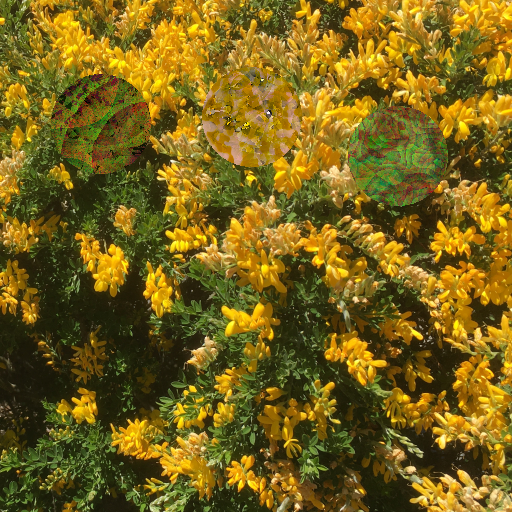}
    \hfill
    \igfour{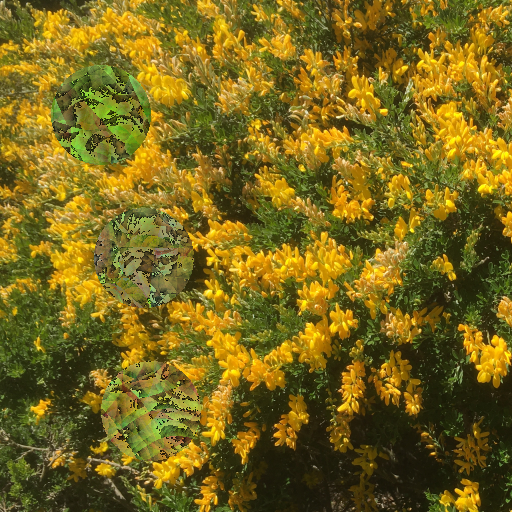}
    \caption{Four tournament images from run {\runID yellow\_flower\_on\_green\_20221217\_1826}. This background set was a notoriously ``hard'' non-stationary test case.}
    \label{fig:yellow_flower_4x}
\end{figure*}

\begin{figure*}[t]
    \igfour{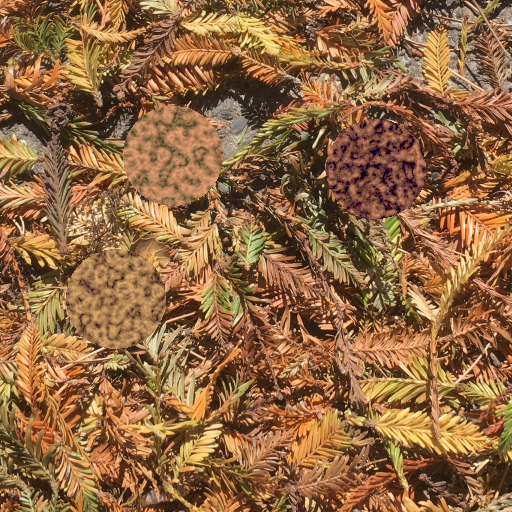}
    \hfill
    \igfour{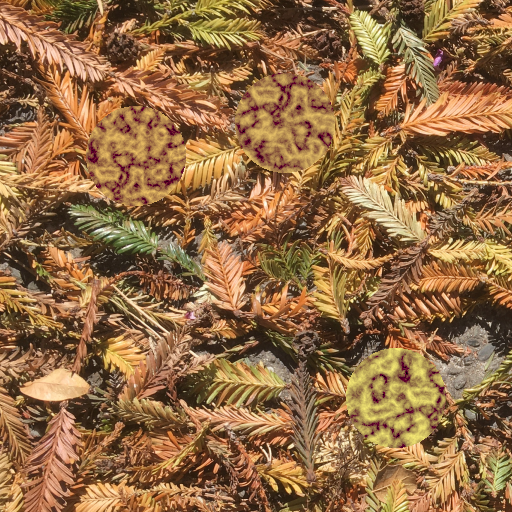}
    \hfill
    \igfour{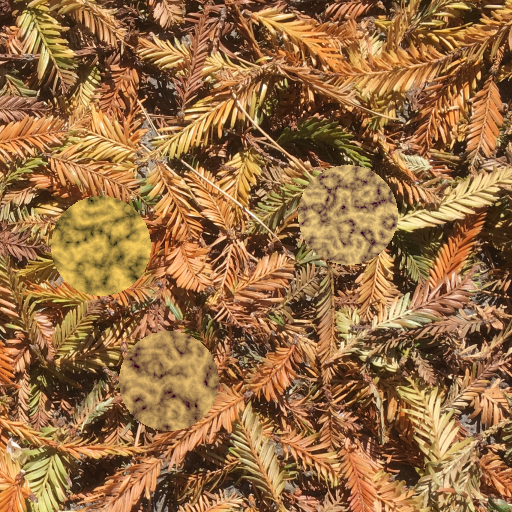}
    \hfill
    \igfour{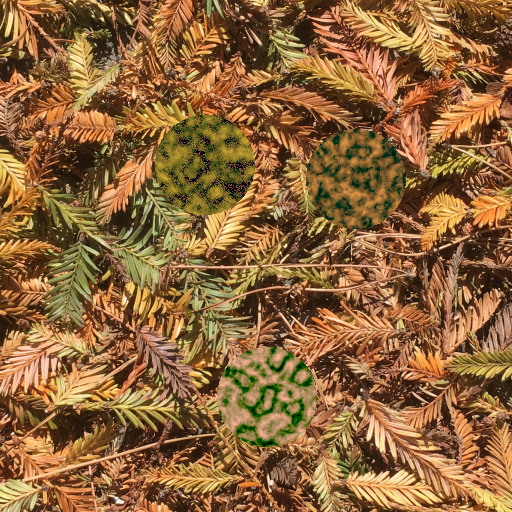}
    \caption{Four tournament images from run {\runID redwood\_leaf\_litter\_20230115\_1730}.}
    \label{fig:redwood_leaf_litter_4x}
\end{figure*}


\subsection{Background Sets}
\label{subsec:background_sets}
Each simulation run is based on a \jargon{background set} of images, usually photographs of natural scenes. These images provide the background of tournament images, over which camouflaged prey are drawn. The images play the role of an \jargon{environment} in which prey must hide to avoid being found and eaten by predators. Because the model is purely 2d, photographs offer an easy way to provide varied environments, of plausible natural complexity, in which to test camouflage evolution.
\par
Background sets used in this work consist of from 1 to 14 photographs, 4 or 5 being typical. Almost all are casual snapshots taken with a mobile phone. The images within a given background set are all similar. For example, a set called \stt{oak\_leaf\_litter} has six images, all of slightly different portions of fallen leaves piled along a roadside. Each image in this set is taken with the camera pointing straight down, all from about the same height. As a result the images have features (e.g. leaves) of about the same size. This ``similarity'' helps the camouflage evolution process by providing many unique yet analogous backgrounds in which to try hiding.
\par
As used here, ``similarity'' is meant to suggest image features have a \jargon{stationary distribution}, that different patches (say of a size comparable to prey) have the same statistical distribution of color and spatial frequency. When background sets are ``less stationary'' (e.g. large areas of uniform color) it becomes \jargon{harder} for evolution to find good camouflage. See faster progress for easy background in Figure \ref{fig:sqm_plot} — the ``hard'' background set shown there is the yellow\_flower\_on\_green set in Figure \ref{fig:yellow_flower_4x}, which includes large regions of mostly yellow, or mostly green, or mostly black.
\par


\begin{figure*}[t]
    \igfour{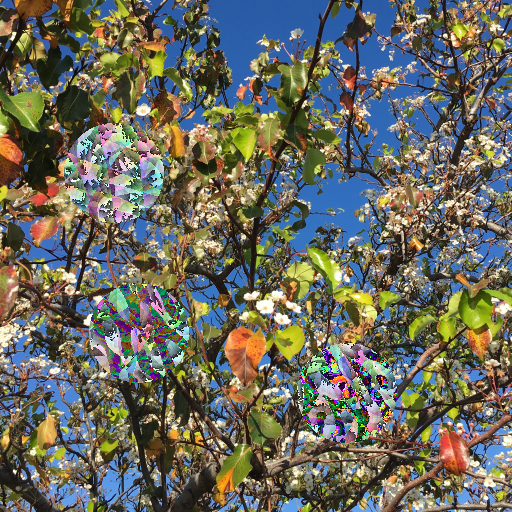}
    \hfill
    \igfour{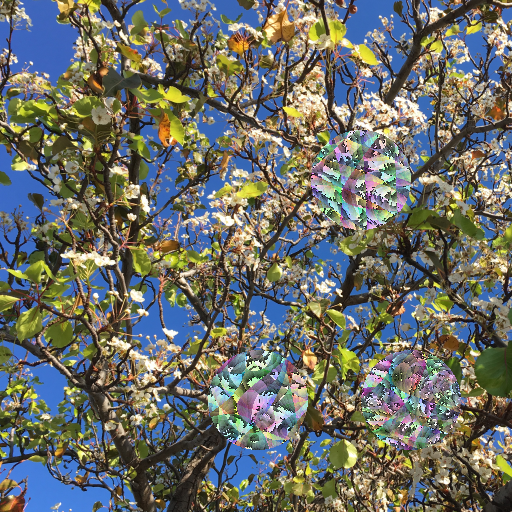}
    \hfill
    \igfour{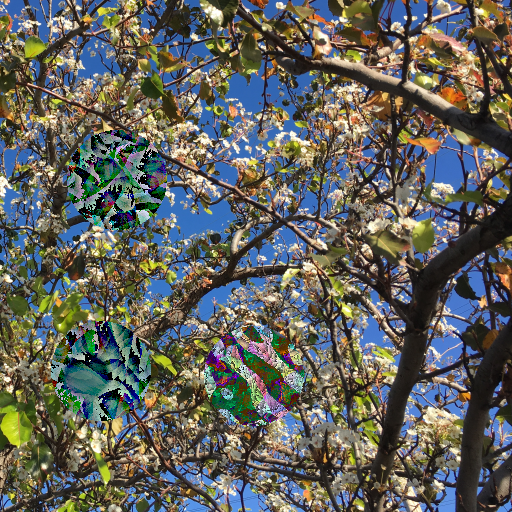}
    \hfill
    \igfour{20221108_2018_step_6562.png}
    \caption{Four tournament images from run {\runID tree\_leaf\_blossom\_sky\_20221108\_2018}.}
    \label{fig:tree_leaf_blossom_sky_4x}
\end{figure*}

\begin{figure*}[t]
    \igfour{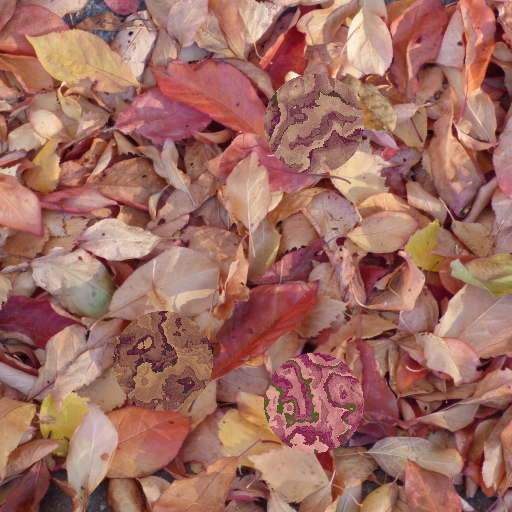}
    \hfill
    \igfour{20221121_1819_step_6464.png}
    \hfill
    \igfour{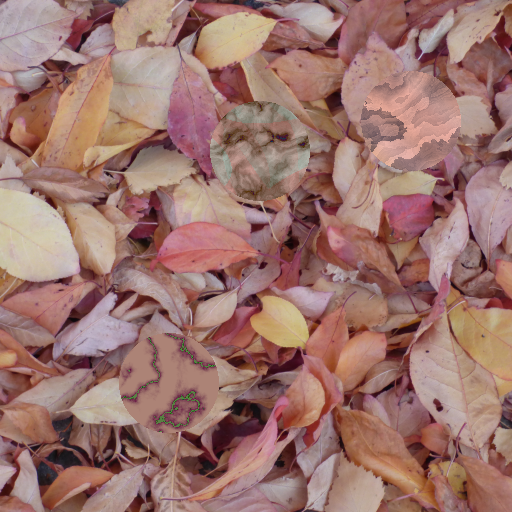}
    \hfill
    \igfour{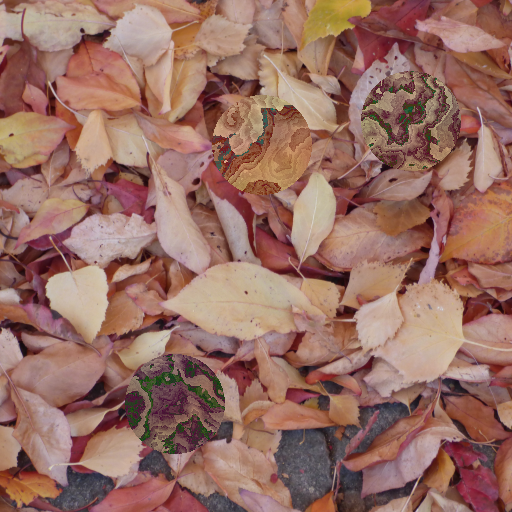}
    \caption{Four tournament images from run {\runID plum\_leaf\_litter\_20221121\_1819}.}
    \label{fig:plum_leaf_litter_4x}
\end{figure*}


\subsection{Simulation Runs}
\label{subsec:simulation_runs}
To run this coevolutionary simulation model, two processes are launched. One is the evolutionary texture synthesis system that models a population of camouflaged prey. It is c++ code based on \texsyn{} and \lazypredator{}. The other is a ``predator server'' that manages a population of visual hunters and fine-tunes their visual perception. This \jargon{\predatoreye{}} \citep{reynolds_predatoreye_2021} is Python code using Keras \citep{chollet_keras_2015} and TensorFlow \citep{tensorflow_whitepaper_2015}. The prey side produces a labeled tournament image. That is inspected by the predator side, which sends back a target location within the image, indicating its estimate of where the most conspicuous prey is located. The labeled tournament image is used to fine-tune the predators. The target location drives the fitness function for prey evolution.
\par
Main parameters for a simulation run are a choice of background set (see Section \nameref{subsec:background_sets}), a scale factor for the background images, and a random number seed. The full set of run parameters are described in Section \nameref{sec:texsyn_cmd_line_args} and Table \ref{table:key_simulation_parameters}.
\par
A typical run consists of 12,000 steps with a prey population of 400 and predator population of 40. A reimplementation of the interactive approach of \citet{reynolds_iec_2011} used a prey population of 120 and typical runs of 2000 to 3000 steps, as did the first version with a neural net predator. Introducing a population of such predators reduced the rate of fine-tuning of predator's neural net models, leading to simulations of 6000, then 12,000 steps.
\par
During a simulation run various data is collected in log files. The most important output is a ``visual log'' --- periodically saving tournament images (a crop of the given backgrounds overlaid with camouflaged prey, as in Figure \ref{fig:teaser}) along with the predator response data. These images are saved ``occasionally.'' Originally it was every 20 steps, but was changed to every 19 steps to be relatively prime and so cycle through the 6, 10, or 20 \jargon{subpopulations} (\jargon{demes} or \jargon{islands}) of the prey population. Eventually, hand selection from this periodic collection was replaced by ``auto-curation'', see Section \nameref{subsec:auto_curation}.
\par


\subsection{Static Quality Metric}
\label{subsec:sqm}

\begin{figure}[t]
    \includegraphics[width=\columnwidth]{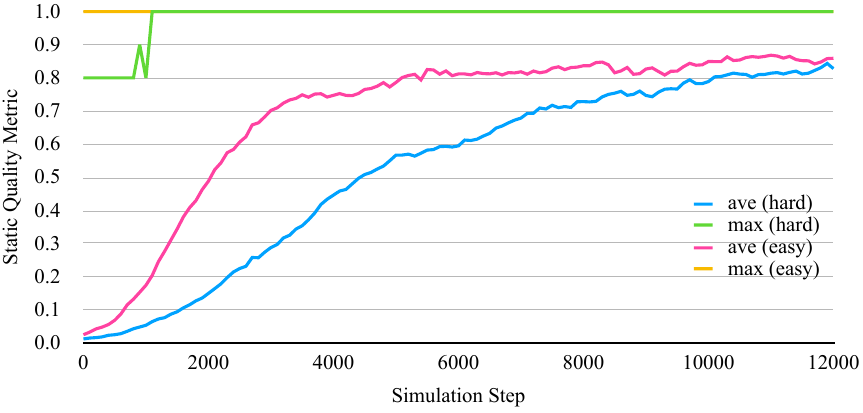}
    \caption{Static quality metric (see Section \nameref{subsec:sqm}) versus simulation time. Metric is based on failure of pre-trained FCD model to find prey. Compares ``easy'' background ({\runID oak\_leaf\_litter}) versus ``hard'' ({\runID yellow\_flower\_on\_green}).}
    \label{fig:sqm_plot}
\end{figure}

\begin{figure}[t]
    \includegraphics[width=\columnwidth]{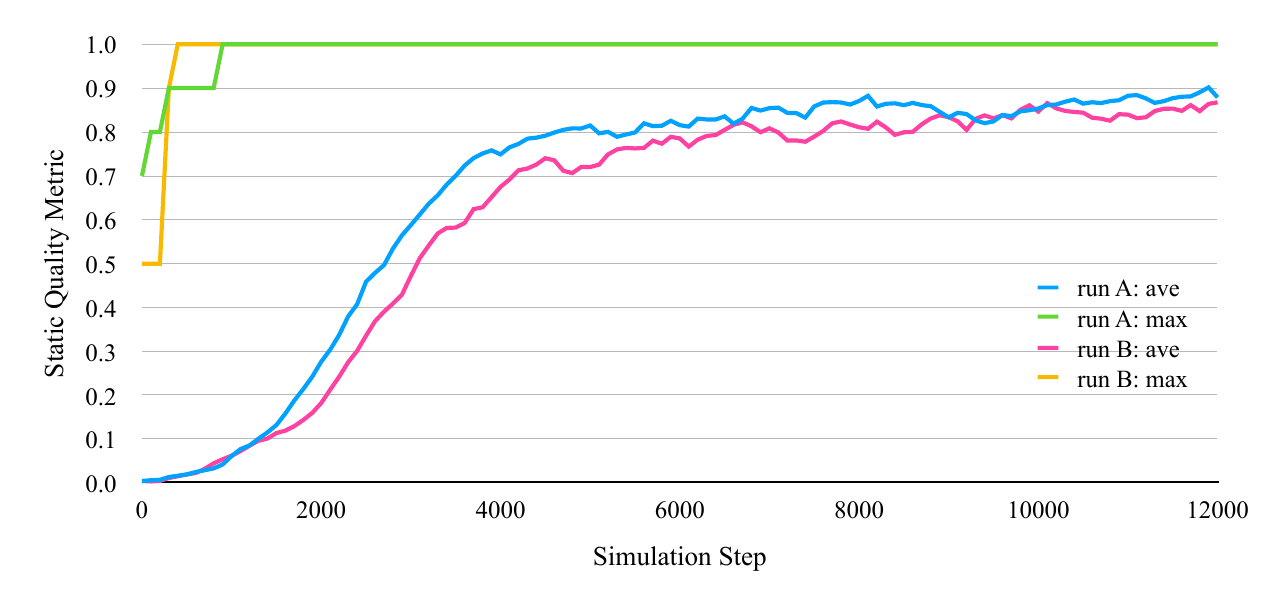}
    \caption{Static quality metric comparing two runs that differ only by the random number seed used. This illustrates the typical variance between simulation runs. Run A ({\runID redwood\_leaf\_litter\_20230319\_1128}) and run B ({\runID redwood\_leaf\_litter\_20230322\_2254}) track along a similar curve during the 12,000 simulation steps. In both cases the population average SQM improves strongly until about step 5000. Then it continues to improve at a much slower rate.}
    \label{fig:sqm_variance}
\end{figure}

To track the objective progress of camouflage coevolution, the pre-trained FCD neural net model can serve as a \jargon{standard predator} to provide an unchanging \jargon{static quality metric}. Each prey is scored by the fraction of ten trials this standard predator is ``fooled'' --- fails to find the prey on a random background. Figure \ref{fig:sqm_plot} uses the SQM to visualize the difference between an easy ``stationary'' background set and a hard ``non-stationary'' background containing large features. Figure \ref{fig:sqm_variance} illustrates the relatively small variance between two runs which were identical except for their random number seed.
\par


\subsection{Automatic Selection of Results from a Run}
\label{subsec:auto_curation}

A typical simulation run produces up to 600 files containing ``tournament images.'' That is a lot to store. Worse, it is a lot to sort through by hand to find a few good quality ``representative'' images, say for publication. This was the procedure used for most of the camouflage images in this paper.
\par
This project has progressed from a human-in-the-loop model to less and less human intervention. In \citet{reynolds_iec_2011} the predator \textit{was} a human. Then a self-organizing predator model was incorporated \citep{reynolds_coevolution_2023}. Finally an \jargon{auto-curation} facility was added to automatically select results from a run. This is similar in spirit to an issue in multi-objective evolutionary algorithms (MOEA): selecting the best from the many solutions clustered along a Pareto optimal boundary (see for example \citet{ishibuchi_difficulties_2022}).
\par
This auto-curation selects candidate images by an objective method, based on the static quality metric (Section \nameref{subsec:sqm}), intended to collect most of the high quality images showing effective camouflage. Nonetheless, the final culling (for publication or application) is likely made by a human.
\par
The current auto-curation filter combines two factors. Selected images must both have a ``perfect'' SQM score, meaning the pre-trained predator fails to find them in any of ten trials. Secondly, images selected by auto-curation must have ``fooled'' (gone undetected by) all three predators in a tournament. The first component is based on ten trials against the pre-trained predator. The second component is based on three trials against fine-tuned predators evolved for this specific background and these specific camouflaged prey. (More details in the September 30, 2023 post in the project blog \citep{reynolds_texsyn_blog_2023})
\par


\section{Discussion}
\label{sec:discussion}
\begin{figure*}
    \igfour{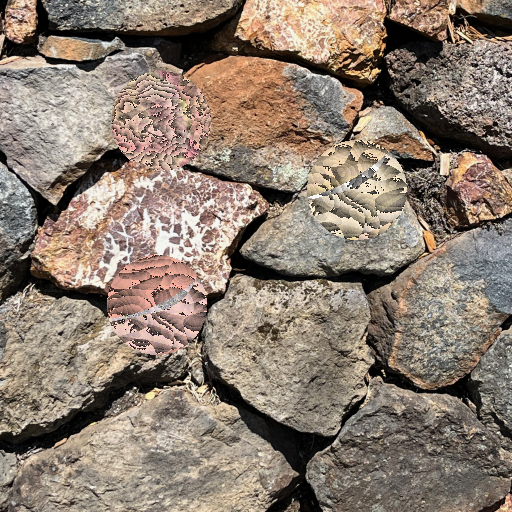}
    \hfill
    \igfour{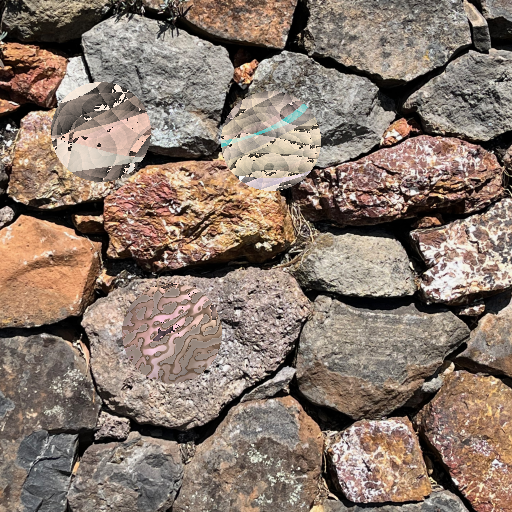}
    \hfill
    \igfour{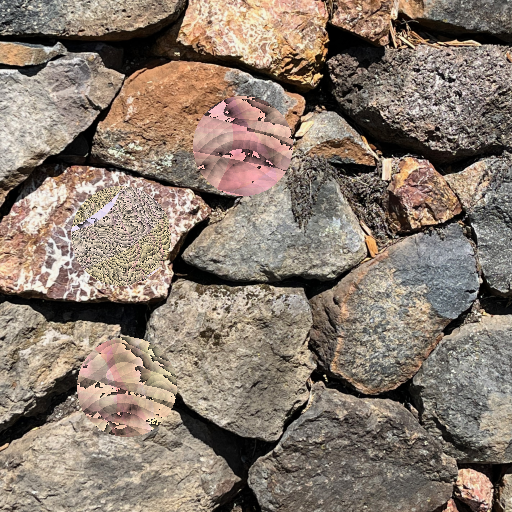}
    \hfill
    \igfour{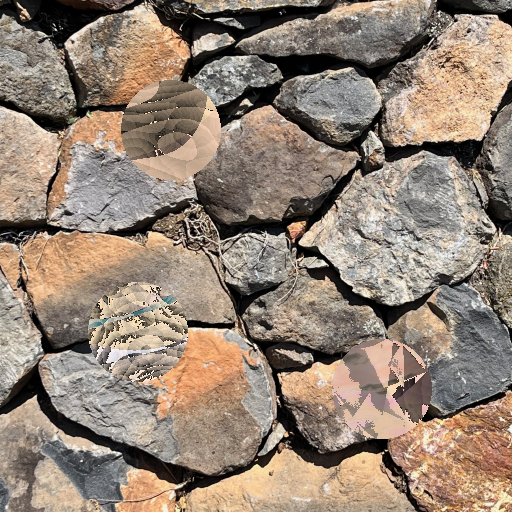}
    \caption{Four tournament images from run {\runID rock\_wall\_20221112\_1555}.}
    \label{fig:rock_wall_4x}
\end{figure*}

This simulation model was applied to a variety of background sets to produce prey camouflage patterns suited to those environments. See examples in Figures \ref{fig:teaser}, \ref{fig:time_sequence}, \ref{fig:rock_wall_4x}, other figures in this paper, and TexSyn's blog \citep{reynolds_texsyn_blog_2023}.
\par
These experiments were run on a 2021 Apple MacBook Pro with an M1 Max chip. Typical simulation runs show the predator vision/learning process taking about ``600\%'' of the processing power (that is, about 6 of 10 cores) and the prey texture render/evolve process taking about 10\% of a core (many steps reuse cached prey renderings). The time to complete a single simulation step is about 1 second. Typical simulation runs are 12,000 steps so take about 3.5 hours. When using the static quality metric to evaluate progress (Figure \ref{fig:sqm_plot}) total simulation run time is about 10 hours.
\par
A key observation from these experiments is that background sets present varying levels of difficulty (Figure \ref{fig:sqm_plot}). For some background sets, a simulated evolution run easily produces effective camouflage. Occasionally it is good enough to momentarily fool a human observer: ``Wait, where is the third prey?!'' Sometimes runs fail to produce effective results. (That is, none of the results appear well-camouflaged to a human observer, or the average SQM for a run never gets above (say) 0.6.) Re-running the simulation with a new random seed will often find a solution. This suggests a ``hard'' background set may have a likelihood of success during a run of say 1/2 or 1/3 while ``easy'' background have likelihood close to 1.
\par
Figure \ref{fig:sqm_variance} shows that these simulations produce  camouflage patterns of similar quality for similar runs. Keeping all simulation parameters constant, except for the initial random number seed, the plot of Static Quality Metric over time is very similar between the two runs. Images of these runs can be seen in the project blog \citep{reynolds_texsyn_blog_2023} entry for March 19, 2023. Note that while those runs have similar SQM scores, visually they are quite dissimilar.
\par


\section{Limitations}
\label{subsec:limitations}
This work has the nature of an ``existence proof.'' The model of camouflage coevolution presented here is a starting point. Its architecture and parameters (Table \ref{table:key_simulation_parameters}) are unlikely to be optimal. It was tuned well enough to allow camouflage evolution to emerge, but surely can be improved through further systematic experimentation.
\par
Another limitation is that all final results are to some extent hand selected: ``cherry picked'' (see for example, Figures \ref{fig:teaser}, \ref{fig:time_sequence}, \ref{fig:rock_wall_4x}). After a simulation run, the resulting images are examined by a human who selects some as representative of the results, by making a mental judgement of their effectiveness.
\par
This cherry-picking is a problem for evaluating the scientific validity of the model: does this technique work well, or is the author just good at cherry-picking effective results? As a practical matter, the \jargon{auto-curation} described in Section \nameref{subsec:auto_curation} allows using this simulation to generate camouflage without human input, and so can avoid potential experimenter bias that might skew results. That ``hands free'' operation enables running large automated processes composed of many camouflage coevolution simulations. For example: making several runs that are identical except for settings of a single parameter.
\par
Other limitations include the inherently 2D nature of the simulation, that simulated time is discrete, and that the model of texture synthesis lacks genetic or biological plausibility. These are among the simplifying abstractions intended to make this initial model of camouflage coevolution tractable. Still it may be fair to complain that as a result, the model is \textit{too} abstract to provide much biological insight.
\par


\begin{figure*}[t]
    \igfour{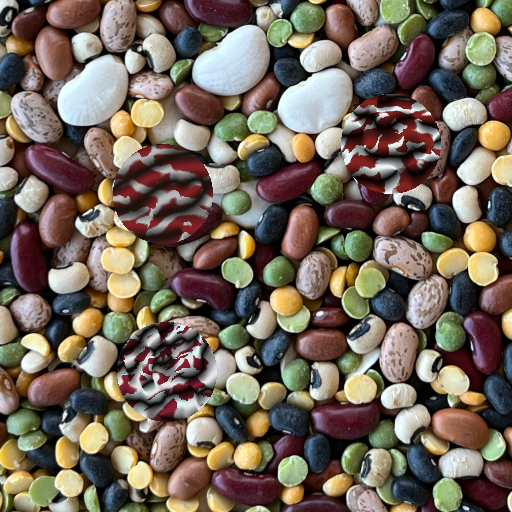}
    \hfill
    \igfour{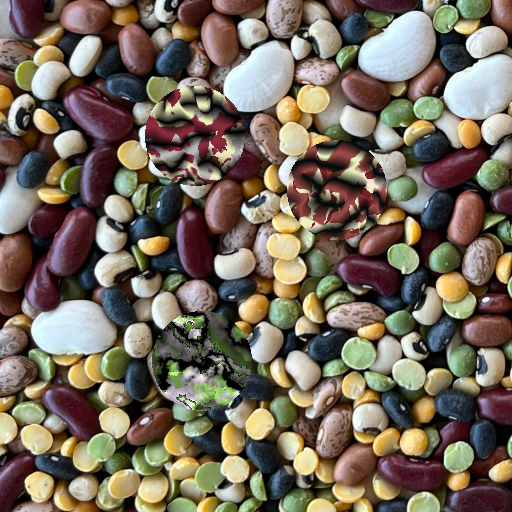}
    \hfill
    \igfour{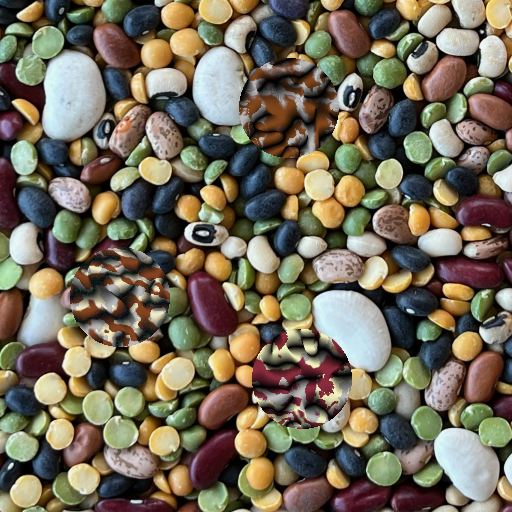}
    \hfill
    \igfour{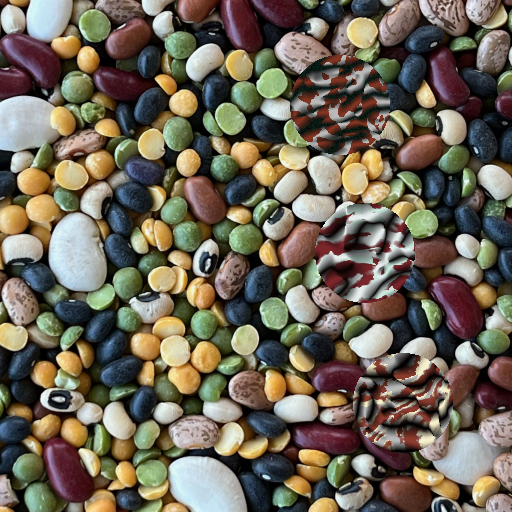}
    \caption{Four tournament images from run {\runID bean\_soup\_mix\_20221231\_1317}.}
    \label{fig:bean_soup_mix_4x}
\end{figure*}

\begin{figure*}[t]
    \igfour{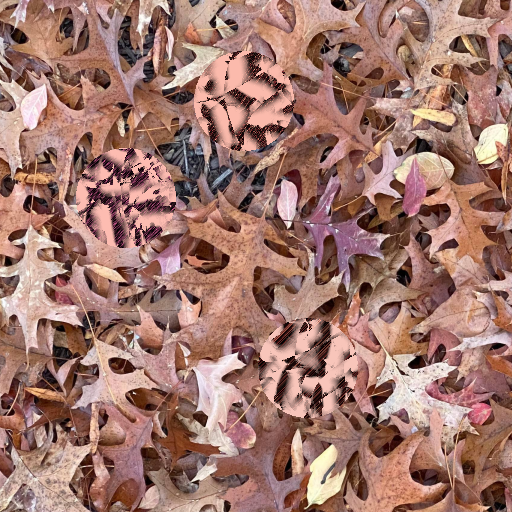}
    \hfill
    \igfour{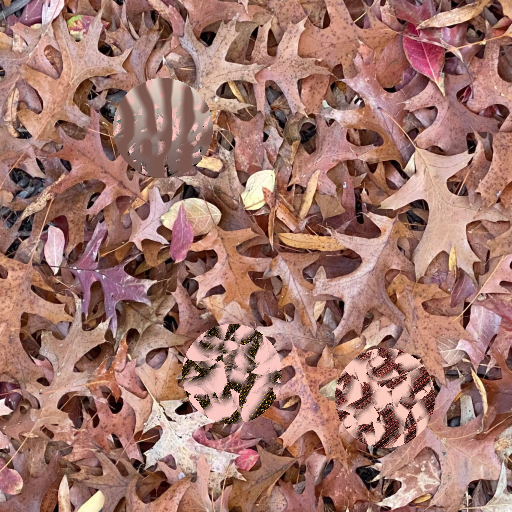}
    \hfill
    \igfour{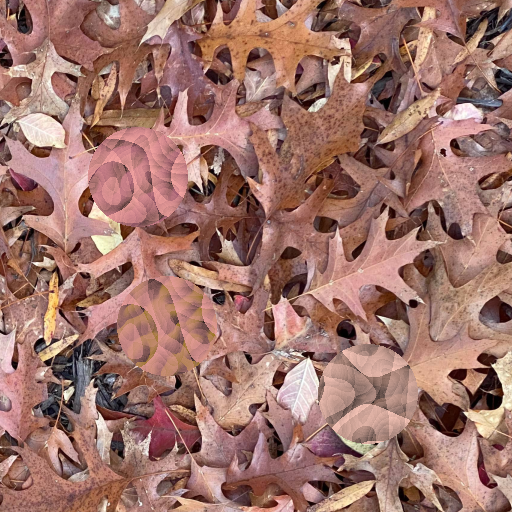}
    \hfill
    \igfour{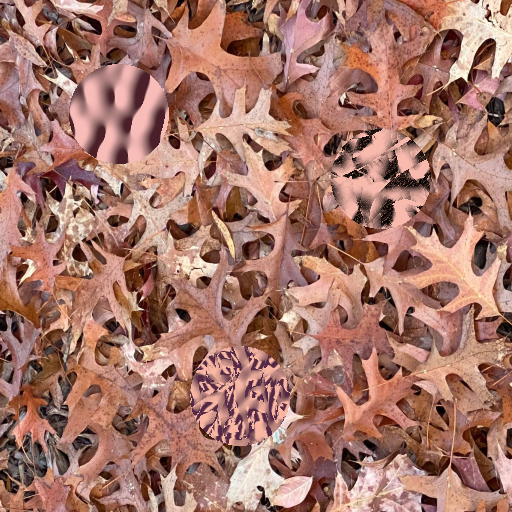}
    \caption{Four tournament images from run {\runID jans\_oak\_leaves\_20221227\_1717}.}
    \label{fig:jans_oak_leaves_4x}
\end{figure*}


\section{Future Work}
The goal of this work has been to build a computational model of the coevolution of camouflage from the interaction of predator and prey. It demonstrates that camouflage can in fact arise from such a system. It also provides a simple process to generate camouflage for a given environment from photos. The most important contribution of the work has been to create an open source model allowing future experiments, to help study camouflage in nature, and the perceptual phenomenon of camouflage more generally.
\par
A \jargon{static quality metric} for camouflage (discussed in Section \nameref{subsec:simulation_runs} and Figures \ref{fig:sqm_plot} and \ref{fig:sqm_variance}) is based on using the pre-trained predator model (see Section \nameref{sec:pre_train_predator}) as a standard. This is not ideal. It seems likely this metric may saturate (reach the top of its range, failing to find a prey in all ten trials) while coevolution continues to improve camouflage quality. Other approaches should be investigated. For example \citet{lv_cod_2022} and \citet{volonakis_camouflage_2018} suggests potential ways to rank ``conspicuousness.'' The model presented here would be a useful testbed for evaluating candidate camouflage metrics. It should also be possible to validate these candidate metrics with crowd-sourced rating of camouflage quality, as in Sensory Ecology and Evolution: Games \citep{stevens_games_2022} and CamoEvo \citep{hancock_camoevo_2022}.
\par
As mentioned in Section \nameref{subsec:background_sets}, and Figure \ref{fig:sqm_plot}, the ``difficulty'' of evolving an effective camouflage pattern for a given background set, seems to be strongly related the the \jargon{stationarity} of the images. Applying information theory metrics for stationarity \citep{conni_visual_2021} to background images beforehand might provide useful predictions of the difficulty of a given background set. This could in turn be used to automatically set \jargon{hyper-parameters} for the run, such as evolutionary population and the number of evolution steps to take.
\par
All experiments described here use a fixed background environment, uniformly sampled from a pre-specified set of background images. This model could easily be be extended to schedule changes of the background images over evolutionary time. These might reflect seasonal changes, longer term climate changes, or animals moving from one ecosystem to another. Not only would this open up many new experimental directions, but as described in \citet{kashtan_varying_2007}: ``varying environments can speed up evolution.''
\par
All of the simulation parameters in Table \ref{table:key_simulation_parameters} should be reexamined using a static camouflage metric. Similarly, the neural net architecture in Figure \ref{fig:predator_cnn} is simply the first one that worked. Other designs should be constructed and evaluated.
\par
This work used a fixed predator model as a static quality metric to evaluate and track changes in prey camouflage quality over time (see Section \nameref{subsec:sqm}). Perhaps it would be useful to do the effectively the opposite, and so provide a way to track \textbf{predator} quality over time. Perhaps a subset of FCD6 \citep{reynolds_FCD6_2022} training images (see Figure \ref{fig:fcd5_examples}), or a random sampling of tournament images from the current run, could be selected as a fixed reference for predator quality. The metric might be the fraction of such test images a predator locates successfully. This sort of metric might help, for example, to visualize the coevolutionary dynamics at work in this simulation.
\par
In early versions, simulated predators often incorrectly predicted prey to be at the tournament image's center. Perhaps this \jargon{center preference} is a ``lazy'' default strategy, since that is the mean over all prey positions. To work around this, \texsyn{}'s random placement was constrained to avoid prey center positions within one prey diameter of the image center, thus forcing predators to hunt prey away from the center. This seems like a bug that should be better understood and fixed in a more principled way.
\par
This simulation is currently a mixed paradigm model using both evolution and learning. While these typically co-occur in nature \citep{valiant_probably_2013}, what about an all-evolution model with evolved detectors for predators? Perhaps along the lines of \citet{harrington_coevolution_2014} and \citet{bi_genetic_2022}. Recent work on evolutionary algorithms for \jargon{merging} deep neural net models \citep{akiba_evolutionary_2024} might provide a useful way to evolve a population of CNN predators. Or conversely, an all learning model, something similar to CamoGAN \citep{talas_camogan_2020} may provide new research directions.
\par
An obvious next step is to apply this model to 3d environments. Perhaps as described in \citet{miller_color_2022}. One key simplifying assumption of the current purely 2d model is that plausible, naturally complex, environments are provided by simple photographs from the phone in our pocket. Providing plausibly complex 3d environments is much harder. Perhaps neural techniques (like NeRF \citep{gao_nerf_2022} or NIP \citep{sharp_spelunking_2022}) would meet that need.
\par



\section{Acknowledgements}
I deeply appreciate everyone who helped me with this work: my family for loving support, Ken Perlin for (well, lots, but especially) PSE \citep{perlin_image_1985}, Andrew Glassner for teaching me everything I know about deep learning \citep{glassner_deep_2021}, and Pat Hanrahan for some key career advice (''just do the research'').
\par
I've been working on this project on and off since 2007, based on inspirations by papers in the early 1990s: \citet{witkin_reaction_1991}, \citet{turk_generating_1991}, \citet{angeline_competitive_1993}, \citet{sims_artificial_1991}, and \citet{sims_evolving_1994}). Also one paper published the year before I was born: \citet{turing_chemical_1952}.
I also wish to thank three sets of reviewers for many helpful suggestions.
Thanks for additional help from:
Bilal Abbasi,
Jan Allbeck,
Rebecca Allen,
Richard Dawkins,
Steve DiPaola,
Aaron Hertzmann,
Bjoern Knafla,
John Koza,
Dominic Mallinson,
Nick Porcino,
Michael Wahrman,
and Lance Williams.
Thanks also to my neighbors whose landscaping provided many of the background images used here, collected on daily walks during COVID-19 lockdown.
\par



\bibliographystyle{apalike}
\bibliography{coc.bib}

\begin{thebibliography}{}

\bibitem[Abadi et~al., 2015]{tensorflow_whitepaper_2015}
Abadi, M. et~al. (2015).
\newblock {TensorFlow}: Large-scale machine learning on heterogeneous systems.
\newblock \url{https://www.tensorflow.org/}.

\bibitem[Akiba et~al., 2024]{akiba_evolutionary_2024}
Akiba, T., Shing, M., Tang, Y., Sun, Q., and Ha, D. (2024).
\newblock Evolutionary {Optimization} of {Model} {Merging} {Recipes}.
\newblock arXiv:2403.13187 [cs].

\bibitem[Angeline and Pollack, 1993]{angeline_competitive_1993}
Angeline, P.~J. and Pollack, J.~B. (1993).
\newblock Competitive {Environments} {Evolve} {Better} {Solutions} for {Complex} {Tasks}.
\newblock In Grefenstette, J.~J., editor, {\em Proceedings {Of} {The} {First} {International} {Conference} {On} {Genetic} {Algorithms} {And} {Their} {Applications}}, pages 1--7, Pittsburgh, PA. Psychology Press.

\bibitem[Bi et~al., 2022]{bi_genetic_2022}
Bi, Y., Xue, B., and Zhang, M. (2022).
\newblock Genetic programming-based evolutionary deep learning for data-efficient image classification.
\newblock {\em IEEE Transactions on Evolutionary Computation}, 0(0):15.

\bibitem[Bradley and Blossom, 2023]{bradley_generation_2023}
Bradley, J.~R. and Blossom, A.~P. (2023).
\newblock The {Generation} of {Visually} {Credible} {Adversarial} {Examples} with {Genetic} {Algorithms}.
\newblock {\em ACM Transactions on Evolutionary Learning and Optimization}, 3(1):2:1--2:44.

\bibitem[Brichard et~al., 2023]{brichard_natural_2023}
Brichard, Y.~H., Raymond, M., Cuthill, I.~C., Mendelson, T.~C., and Renoult, J.~P. (2023).
\newblock From natural to sexual selection: {Revealing} a hidden preference for camouflage patterns.
\newblock Pages: 2023.09.27.559753 Section: New Results.

\bibitem[Casgrain et~al., 2022]{casgrain_deep_2022}
Casgrain, P., Ning, B., and Jaimungal, S. (2022).
\newblock Deep {Q}-{Learning} for {Nash} {Equilibria}: {Nash}-{DQN}.
\newblock arXiv:1904.10554 [cs, q-fin, stat].

\bibitem[Chastain et~al., 2013]{chastain_multiplicative_2013}
Chastain, E., Livnat, A., Papadimitriou, C., and Vazirani, U. (2013).
\newblock Multiplicative updates in coordination games and the theory of evolution.
\newblock In {\em Proceedings of the 4th conference on {Innovations} in {Theoretical} {Computer} {Science}}, pages 57--58. Association for Computing Machinery, New York, NY, USA.

\bibitem[Chen et~al., 2022]{chen_boundary-guided_2022}
Chen, T., Xiao, J., Hu, X., Zhang, G., and Wang, S. (2022).
\newblock Boundary-guided network for camouflaged object detection.
\newblock {\em Knowledge-Based Systems}, 248:108901.

\bibitem[Chen et~al., 2023a]{chen_diffusion_2023}
Chen, Z., Gao, R., Xiang, T.-Z., and Lin, F. (2023a).
\newblock Diffusion {Model} for {Camouflaged} {Object} {Detection}.
\newblock arXiv:2308.00303 [cs].

\bibitem[Chen et~al., 2023b]{chen_camodiffusion_2023}
Chen, Z., Sun, K., Lin, X., and Ji, R. (2023b).
\newblock {CamoDiffusion}: {Camouflaged} {Object} {Detection} via {Conditional} {Diffusion} {Models}.
\newblock arXiv:2305.17932 [cs].

\bibitem[Chollet et~al., 2015]{chollet_keras_2015}
Chollet, F. et~al. (2015).
\newblock Keras.
\newblock \url{https://keras.io}.

\bibitem[Chu et~al., 2010]{chu_camo_image_2010}
Chu, H.-K., Hsu, W.-H., Mitra, N.~J., Cohen-Or, D., Wong, T.-T., and Lee, T.-Y. (2010).
\newblock Camouflage images.
\newblock {\em ACM Trans. Graph.}, 29(4).

\bibitem[Conni et~al., 2021]{conni_visual_2021}
Conni, M., Deborah, H., Nussbaum, P., and Green, P. (2021).
\newblock Visual and data stationarity of texture images.
\newblock {\em Journal of Electronic Imaging}, 30(4):043001.
\newblock Publisher: SPIE.

\bibitem[Cott, 1940]{cott_adaptive_1940}
Cott, H.~B. (1940).
\newblock {\em Adaptive {Coloration} in {Animals}}.
\newblock Methuen and Co, London.

\bibitem[Cramer, 1985]{cramer_representation_1985}
Cramer, N.~L. (1985).
\newblock A {Representation} for the {Adaptive} {Generation} of {Simple} {Sequential} {Programs}.
\newblock In {\em Proceedings of an {International} {Conference} {On} {Genetic} {Algorithms} {And} {Their} {Applications}}, pages 183--187, Carnegie-Mellon University, Pittsburgh, PA, USA. Lawrence Erlbaum Associates.

\bibitem[Cuthill, 2019]{cuthill_camouflage_2019}
Cuthill, I.~C. (2019).
\newblock Camouflage.
\newblock {\em Journal of Zoology}, 308(2):75--92.

\bibitem[Dawkins, 1986]{dawkins_blind_1986}
Dawkins, R. (1986).
\newblock {\em The {Blind} {Watchmaker}}.
\newblock W. W. Norton \& Company, Inc., New York, NY, USA.

\bibitem[de~Alcantara~Viana et~al., 2022]{de_alcantara_viana_predator_2022}
de~Alcantara~Viana, J.~V., Vieira, C., Duarte, R.~C., and Romero, G.~Q. (2022).
\newblock Predator responses to prey camouflage strategies: a meta-analysis.
\newblock {\em Proceedings of the Royal Society B: Biological Sciences}, 289(1982):20220980.
\newblock Publisher: Royal Society.

\bibitem[De~Gomensoro~Malheiros et~al., 2020]{de_gomensoro_malheiros_leopard_2020}
De~Gomensoro~Malheiros, M., Fensterseifer, H., and Walter, M. (2020).
\newblock The leopard never changes its spots: realistic pigmentation pattern formation by coupling tissue growth with reaction-diffusion.
\newblock {\em ACM Transactions on Graphics}, 39(4):63:63:1--63:62:14.

\bibitem[Endler, 1978]{endler_predators_1978}
Endler, J.~A. (1978).
\newblock A {Predator}’s {View} of {Animal} {Color} {Patterns}.
\newblock In Hecht, M.~K., Steere, W.~C., and Wallace, B., editors, {\em Evolutionary {Biology}}, Evolutionary {Biology}, pages 319--364. Springer US, Boston, MA.

\bibitem[Endler, 1980]{endler_natural_1980}
Endler, J.~A. (1980).
\newblock Natural {Selection} on {Color} {Patterns} in {Poecilia} reticulata.
\newblock {\em Evolution}, 34(1):76--91.
\newblock Publisher: [Society for the Study of Evolution, Wiley].

\bibitem[Endler, 2012]{endler_framework_2012}
Endler, J.~A. (2012).
\newblock A framework for analysing colour pattern geometry: adjacent colours.
\newblock {\em Biological Journal of the Linnean Society}, 107(2):233--253.

\bibitem[Gao et~al., 2022]{gao_nerf_2022}
Gao, K., Gao, Y., He, H., Lu, D., Xu, L., and Li, J. (2022).
\newblock {NeRF}: {Neural} {Radiance} {Field} in {3D} {Vision}, {A} {Comprehensive} {Review}.
\newblock \url{http://arxiv.org/abs/2210.00379}.
\newblock arXiv:2210.00379 [cs].

\bibitem[Glassner, 2021]{glassner_deep_2021}
Glassner, A. (2021).
\newblock {\em Deep {Learning}: {A} {Visual} {Approach}}.
\newblock No Starch Press, San Francisco, CA.

\bibitem[Goodfellow et~al., 2014]{goodfellow_gan_2014}
Goodfellow, I.~J., Pouget-Abadie, J., Mirza, M., Xu, B., Warde-Farley, D., Ozair, S., Courville, A., and Bengio, Y. (2014).
\newblock Generative adversarial networks.
\newblock \url{https://arxiv.org/abs/1406.2661}.
\newblock Republished in 2020 in CACM: https://doi.org/10.1145/3422622.

\bibitem[Guerrero et~al., 2022]{Guerrero_MatFormer_2022}
Guerrero, P., Ha\v{s}an, M., Sunkavalli, K., M\v{e}ch, R., Boubekeur, T., and Mitra, N.~J. (2022).
\newblock Matformer: A generative model for procedural materials.
\newblock {\em ACM Trans. Graph.}, 41(4).

\bibitem[Guo et~al., 2022]{guo_ganmouflage_2022}
Guo, R., Collins, J., de~Lima, O., and Owens, A. (2022).
\newblock {GANmouflage}: {3D} object nondetection with texture fields.
\newblock \url{http://arxiv.org/abs/2201.07202}.
\newblock arXiv: 2201.07202.

\bibitem[Hancock and Troscianko, 2022]{hancock_camoevo_2022}
Hancock, G. R.~A. and Troscianko, J. (2022).
\newblock Camoevo: An open access toolbox for artificial camouflage evolution experiments.
\newblock {\em Evolution}, 76(5):870--882.

\bibitem[Harrington et~al., 2014]{harrington_coevolution_2014}
Harrington, K.~I., Freeman, J., and Pollack, J. (2014).
\newblock {Coevolution in Hide and Seek: Camouflage and Vision}.
\newblock In {\em ALIFE 14: The Fourteenth International Conference on the Synthesis and Simulation of Living Systems}, pages 25--32, New York City, NY, USA. MIT Press.

\bibitem[Holland, 1984]{holland_genetic_1984}
Holland, J.~H. (1984).
\newblock Genetic {Algorithms} and {Adaptation}.
\newblock In Selfridge, O.~G., Rissland, E.~L., and Arbib, M.~A., editors, {\em Adaptive {Control} of {Ill}-{Defined} {Systems}}, pages 317--333. Springer US, Boston, MA.

\bibitem[Hu et~al., 2024]{hu_shifting_2024}
Hu, Y., Zhang, J., Zhang, K., and Yuan, Y. (2024).
\newblock Shifting {Spotlight} for {Co}-supervision: {A} {Simple} yet {Efficient} {Single}-branch {Network} to {See} {Through} {Camouflage}.
\newblock arXiv:2404.08936 [cs].

\bibitem[Hughes et~al., 2019]{hughes_imperfect_2019}
Hughes, A., Liggins, E., and Stevens, M. (2019).
\newblock Imperfect camouflage: how to hide in a variable world?
\newblock {\em Proceedings of the Royal Society B: Biological Sciences}, 286(1902):20190646.
\newblock Publisher: Royal Society.

\bibitem[Ishibuchi et~al., 2022]{ishibuchi_difficulties_2022}
Ishibuchi, H., Pang, L.~M., and Shang, K. (2022).
\newblock Difficulties in {Fair} {Performance} {Comparison} of {Multi}-{Objective} {Evolutionary} {Algorithms}.
\newblock {\em IEEE Computational Intelligence Magazine}, 17(1):86--101.

\bibitem[Jaśkowski et~al., 2008]{jaskowski_fitnessless_2008}
Jaśkowski, W., Krawiec, K., and Wieloch, B. (2008).
\newblock Fitnessless coevolution.
\newblock In {\em Proceedings of the 10th annual conference on {Genetic} and evolutionary computation}, {GECCO} '08, pages 355--362, New York, NY, USA. Association for Computing Machinery.

\bibitem[Kashtan et~al., 2007]{kashtan_varying_2007}
Kashtan, N., Noor, E., and Alon, U. (2007).
\newblock Varying environments can speed up evolution.
\newblock {\em Proceedings of the National Academy of Sciences}, 104(34):13711--13716.

\bibitem[Kelley et~al., 2023]{kelley_role_2023}
Kelley, J.~L., Jessop, A.-L., Kelley, L.~A., and Troscianko, J. (2023).
\newblock The role of pictorial cues and contrast for camouflage.
\newblock {\em Evolutionary Ecology}.

\bibitem[Kimura, 1968]{kimura_evolutionary_1968}
Kimura, M. (1968).
\newblock Evolutionary rate at the molecular level.
\newblock {\em Nature insight : aids.}, 217(129):624--626.

\bibitem[Koza, 1992]{koza_genetic_1992}
Koza, J.~R. (1992).
\newblock {\em Genetic {Programming}: {On} the {Programming} of {Computers} by {Means} of {Natural} {Selection} ({Complex} {Adaptive} {Systems})}.
\newblock MIT Press, A Bradford Book, Cambridge, Mass., 1 edition.

\bibitem[Latham, 1989]{latham_form_1989}
Latham, W. (1989).
\newblock {\em Form Synth: The Rule-Based Evolution of Complex Forms from Geometric Primitives}, page 80–108.
\newblock Springer-Verlag, Berlin, Heidelberg.

\bibitem[Livnat and Papadimitriou, 2016]{livnat_sex_2016}
Livnat, A. and Papadimitriou, C. (2016).
\newblock Sex {As} an {Algorithm}: {The} {Theory} of {Evolution} {Under} the {Lens} of {Computation}.
\newblock {\em Commun. ACM}, 59(11):84--93.

\bibitem[Lv et~al., 2022]{lv_cod_2022}
Lv, Y., Zhang, J., Dai, Y., Li, A., Barnes, N., and Fan, D.-P. (2022).
\newblock Towards deeper understanding of camouflaged object detection.
\newblock \url{https://arxiv.org/abs/2205.11333}.

\bibitem[Mckay et~al., 2010]{Mckay_2010}
Mckay, R.~I., Hoai, N.~X., Whigham, P.~A., Shan, Y., and O'Neill, M. (2010).
\newblock Grammar-based genetic programming: A survey.
\newblock {\em Genetic Programming and Evolvable Machines}, 11(3–4):365–396.

\bibitem[Miller et~al., 2022]{miller_color_2022}
Miller, A.~E., Hogan, B.~G., and Stoddard, M.~C. (2022).
\newblock Color in motion: {Generating} 3-dimensional multispectral models to study dynamic visual signals in animals.
\newblock {\em Frontiers in Ecology and Evolution}, 10.

\bibitem[Montana, 1995]{montana_strongly_1995}
Montana, D.~J. (1995).
\newblock Strongly {Typed} {Genetic} {Programming}.
\newblock {\em Evolutionary Computation}, 3(2):199--230.

\bibitem[Murray, 1988]{murray_how_1988}
Murray, J.~D. (1988).
\newblock How the leopard gets its spots.
\newblock {\em Scientific American}, 258(3):80--87.

\bibitem[Nguyen et~al., 2023]{nguyen_few-shot_2023}
Nguyen, T.-D., Vu, A.-K.~N., Nguyen, N.-D., Nguyen, V.-T., Ngo, T.~D., Do, T.-T., Tran, M.-T., and Nguyen, T.~V. (2023).
\newblock Few-shot {Camouflaged} {Animal} {Detection} and {Segmentation}.
\newblock arXiv:2304.07444 [cs].

\bibitem[Owens et~al., 2014]{owens_camouflaging_2014}
Owens, A., Barnes, C., Flint, A., Singh, H., and Freeman, W. (2014).
\newblock Camouflaging an object from many viewpoints.
\newblock In {\em 2014 IEEE Conference on Computer Vision and Pattern Recognition}, pages 2782--2789, Columbus, OH, USA. IEEE CVPR.

\bibitem[Pang et~al., 2022]{pang_zoom_2022}
Pang, Y., Zhao, X., Xiang, T.-Z., Zhang, L., and Lu, H. (2022).
\newblock Zoom {In} and {Out}: {A} {Mixed}-scale {Triplet} {Network} for {Camouflaged} {Object} {Detection}.
\newblock In {\em 2022 {IEEE}/{CVF} {Conference} on {Computer} {Vision} and {Pattern} {Recognition} ({CVPR})}, pages 2150--2160, New Orleans, LA, USA. IEEE.

\bibitem[Perlin, 1985]{perlin_image_1985}
Perlin, K. (1985).
\newblock An image synthesizer.
\newblock {\em SIGGRAPH '85: Proceedings of the 12th annual conference on Computer graphics and interactive techniques}, 19(3):287--296.

\bibitem[Price et~al., 2019]{price_background_2019}
Price, N., Green, S., Troscianko, J., Tregenza, T., and Stevens, M. (2019).
\newblock Background matching and disruptive coloration as habitat-specific strategies for camouflage.
\newblock {\em Scientific Reports}, 9(1):7840.
\newblock Publisher: Nature Publishing Group.

\bibitem[Reynolds, 2011]{reynolds_iec_2011}
Reynolds, C. (2011).
\newblock {Interactive Evolution of Camouflage}.
\newblock {\em Artificial Life}, 17(2):123--136.

\bibitem[Reynolds, 2019]{reynolds_texsyn_2019}
Reynolds, C. (2019).
\newblock {TexSyn}: Library for evolutionary texture synthesis.
\newblock \url{https://github.com/cwreynolds/TexSyn}.

\bibitem[Reynolds, 2021]{reynolds_predatoreye_2021}
Reynolds, C. (2021).
\newblock {PredatorEye}: anti-camouflage, camouflage-breaking.
\newblock \url{https://github.com/cwreynolds/PredatorEye}.

\bibitem[Reynolds, 2022a]{reynolds_FCD6_2022}
Reynolds, C. (2022a).
\newblock {FCD6} {({Find} {Conspicuous} {Disk} {CNN} {Model:} {Version} 6 (rc4))}.

\bibitem[Reynolds, 2022b]{reynolds_lazypredator_2020}
Reynolds, C. (2022b).
\newblock {LazyPredator}: Genetic programming, negative selection, genetic drift.
\newblock \url{https://github.com/cwreynolds/LazyPredator}.

\bibitem[Reynolds, 2023a]{reynolds_coevolution_2023}
Reynolds, C. (2023a).
\newblock Coevolution of {Camouflage}.
\newblock In {\em ALIFE 2023: Ghost in the Machine: Proceedings of the 2023 Artificial Life Conference}. MIT Press.

\bibitem[Reynolds, 2023b]{reynolds_texsyn_blog_2023}
Reynolds, C. (2023b).
\newblock {TexSyn} {Development} {Blog}.
\newblock \url{https://cwreynolds.github.io/TexSyn/}.

\bibitem[Romera-Paredes et~al., 2023]{romera-paredes_mathematical_2023}
Romera-Paredes, B., Barekatain, M., Novikov, A., Balog, M., Kumar, M.~P., Dupont, E., Ruiz, F. J.~R., Ellenberg, J.~S., Wang, P., Fawzi, O., Kohli, P., and Fawzi, A. (2023).
\newblock Mathematical discoveries from program search with large language models.
\newblock {\em Nature}, pages 1--3.
\newblock Publisher: Nature Publishing Group.

\bibitem[Sharp and Jacobson, 2022]{sharp_spelunking_2022}
Sharp, N. and Jacobson, A. (2022).
\newblock Spelunking the deep: Guaranteed queries on general neural implicit surfaces via range analysis.
\newblock {\em ACM Trans. Graph.}, 41(4).

\bibitem[Sims, 1991]{sims_artificial_1991}
Sims, K. (1991).
\newblock Artificial evolution for computer graphics.
\newblock In {\em {SIGGRAPH} '91: {Proceedings} of the 18th annual conference on {Computer} graphics and interactive techniques}, volume~25, pages 319--328, New York, NY, USA. ACM.

\bibitem[Sims, 1994]{sims_evolving_1994}
Sims, K. (1994).
\newblock Evolving 3d morphology and behavior by competition.
\newblock {\em Artif. Life}, 1(4):353–372.

\bibitem[Song et~al., 2023]{song_camouflaged_2023}
Song, Y., Li, X., and Qi, L. (2023).
\newblock Camouflaged {Object} {Detection} with {Feature} {Grafting} and {Distractor} {Aware}.
\newblock arXiv:2307.03943 [cs].

\bibitem[Stevens et~al., 2022]{stevens_games_2022}
Stevens, M. et~al. (2022).
\newblock Sensory {Ecology} and {Evolution}: Games. [{Archived} web page.].

\bibitem[Sun et~al., 2022]{sun_boundary-guided_2022}
Sun, Y., Wang, S., Chen, C., and Xiang, T.-Z. (2022).
\newblock Boundary-{Guided} {Camouflaged} {Object} {Detection}.
\newblock \url{http://arxiv.org/abs/2207.00794}.
\newblock arXiv:2207.00794 [cs] type: article.

\bibitem[Syswerda, 1991]{syswerda_study_1991}
Syswerda, G. (1991).
\newblock A {Study} of {Reproduction} in {Generational} and {Steady}-{State} {Genetic} {Algorithms}.
\newblock In Rawlins, G. J.~E., editor, {\em Foundations of {Genetic} {Algorithms}}, volume~1, pages 94--101. Elsevier, Amsterdam.

\bibitem[Talas et~al., 2020]{talas_camogan_2020}
Talas, L., Fennell, J.~G., Kjernsmo, K., Cuthill, I.~C., Scott-Samuel, N.~E., and Baddeley, R.~J. (2020).
\newblock Camogan: Evolving optimum camouflage with generative adversarial networks.
\newblock {\em Methods in Ecology and Evolution}, 11(2):240--247.

\bibitem[Thayer, 1909]{thayer_concealing-coloration_1909}
Thayer, G.~H. (1909).
\newblock {\em Concealing-coloration in the animal kingdom: an exposition of the laws of disguise through color and pattern: being a summary of {Abbott} {H}. {Thayer}'s discoveries.}
\newblock Macmillan, New York, NY.

\bibitem[Todd and Latham, 1994]{todd_evolutionary_1994}
Todd, S. and Latham, W. (1994).
\newblock {\em Evolutionary Art and Computers}.
\newblock Academic Press, Inc., USA.

\bibitem[Troscianko et~al., 2017]{troscianko_quantifying_2017}
Troscianko, J., Skelhorn, J., and Stevens, M. (2017).
\newblock Quantifying camouflage: how to predict detectability from appearance.
\newblock {\em BMC Evolutionary Biology}, 17(1):7.

\bibitem[Turing, 1952]{turing_chemical_1952}
Turing, A.~M. (1952).
\newblock The {Chemical} {Basis} of {Morphogenesis}.
\newblock {\em Philosophical Transactions of the Royal Society of London. Series B, Biological Sciences}, 237(641):37--72.

\bibitem[Turk, 1991]{turk_generating_1991}
Turk, G. (1991).
\newblock Generating textures on arbitrary surfaces using reaction-diffusion.
\newblock {\em SIGGRAPH Comput. Graph.}, 25(4):289–298.

\bibitem[Valiant, 2013]{valiant_probably_2013}
Valiant, L. (2013).
\newblock {\em Probably {Approximately} {Correct}: {Nature}'s {Algorithms} for {Learning} and {Prospering} in a {Complex} {World}}.
\newblock Basic Books, New York.

\bibitem[Van~Valen, 1973]{van_valen_new_1973}
Van~Valen, L. (1973).
\newblock A new evolutionary law.
\newblock {\em Evolutionary Theory}, 1:1--30.

\bibitem[visionxiang, 2022]{visionxiang_cod}
visionxiang (2022).
\newblock Camouflaged/concealed object detection.
\newblock \url{https://github.com/visionxiang/awesome-camouflaged-object-detection}.

\bibitem[Volonakis et~al., 2018]{volonakis_camouflage_2018}
Volonakis, T.~N., Matthews, O.~E., Liggins, E., Baddeley, R.~J., Scott-Samuel, N.~E., and Cuthill, I.~C. (2018).
\newblock Camouflage assessment: {Machine} and human.
\newblock {\em Computers in Industry}, 99:173--182.

\bibitem[Vu et~al., 2023]{vu_leveraging_2023}
Vu, T.-A., Nguyen, D.~T., Guo, Q., Hua, B.-S., Chung, N.~M., Tsang, I.~W., and Yeung, S.-K. (2023).
\newblock Leveraging {Open}-{Vocabulary} {Diffusion} to {Camouflaged} {Instance} {Segmentation}.
\newblock arXiv:2312.17505 [cs].

\bibitem[Witkin and Kass, 1991]{witkin_reaction_1991}
Witkin, A. and Kass, M. (1991).
\newblock Reaction-diffusion textures.
\newblock {\em SIGGRAPH Comput. Graph.}, 25(4):299–308.

\bibitem[Wu and Huang, 2021]{wu_mimicry_2021}
Wu, Z. and Huang, L. (2021).
\newblock Mimicry: {Genetic}-algorithm-based {Real}-time {System} of {Virtual} {Insects} in a {Living} {Environment}-{A} {New} and {Altered} {Nature}.
\newblock {\em Proceedings of the ACM on Computer Graphics and Interactive Techniques}, 4(2):28:1--28:8.

\bibitem[Xiuxia et~al., 2023]{xiuxia_imitation_2023}
Xiuxia, C., Pin, Z., and Shuaibin, D. (2023).
\newblock Imitation camouflage synthesis based on shallow neural network.
\newblock {\em Multimedia Systems}, 29(5):2705--2714.

\bibitem[Yin et~al., 2022]{yin_camoformer_2022}
Yin, B., Zhang, X., Hou, Q., Sun, B.-Y., Fan, D.-P., and Van~Gool, L. (2022).
\newblock Camoformer: Masked separable attention for camouflaged object detection.
\newblock \url{https://arxiv.org/abs/2212.06570}.

\bibitem[Zhang et~al., 2023]{zhang_camouflaged_2023}
Zhang, H., Qin, C., Yin, Y., and Fu, Y. (2023).
\newblock Camouflaged {Image} {Synthesis} {Is} {All} {You} {Need} to {Boost} {Camouflaged} {Detection}.
\newblock arXiv:2308.06701 [cs].

\bibitem[Zhang et~al., 2022]{Zhang2022}
Zhang, M., Xu, S., Piao, Y., Shi, D., Lin, S., and Lu, H. (2022).
\newblock {PreyNet}: Preying on camouflaged objects.
\newblock In {\em Proceedings of the 30th ACM International Conference on Multimedia}, MM '22, page 5323–5332, New York, NY, USA. Association for Computing Machinery.

\bibitem[Zhang et~al., 2020]{Zhang_Yin_Nie_Zheng_2020}
Zhang, Q., Yin, G., Nie, Y., and Zheng, W.-S. (2020).
\newblock Deep camouflage images.
\newblock {\em Proceedings of the AAAI Conference on Artificial Intelligence}, 34(07):12845--12852.

\bibitem[Zhang et~al., 2013]{zhang_spatial_2013}
Zhang, Y., Xue, S.-q., Jiang, X.-j., Mu, J.-y., and Yi, Y. (2013).
\newblock The {Spatial} {Color} {Mixing} {Model} of {Digital} {Camouflage} {Pattern}.
\newblock {\em Defence Technology}, 9(3):157--161.

\end{thebibliography}



\appendix
\onecolumn
\section{Appendix}


\subsection{Key simulation parameters}

\begin{footnotesize}
\begin{table}[h]
    \centering
    \vspace{0.7cm}

    \begin{tabular}{ |l|r|r| }
        \hline
        \textbf{Parameter} & \textbf{value} \\ 
        \hline
        predator population & 40 \\ 
        prey population & 400 \\ 
        prey subpopulations (demes) & 20 \\
        prey max init tree size & 100 \\
        prey min tree size after crossover & 50 \\
        prey max tree size after crossover & 150 \\
        \hline
        prey render diameter (pixels) & 100 \\ 
        tournament output image size & 512$\times$512 \\ 
        predator input image size & 128$\times$128 \\ 
        \hline
        simulation steps per run (typical) & 12,000 \\
        prey generations equiv (steps/pop) & 30 \\
        predator fail rate (typical) & 15\%-30\% \\
        predator starvation threshold & \\
        \hspace{0.2cm}(success in previous 20 attempts) & $<$40\% \\ 
        \hline
        predator ``FCD'' pre-training: & \\
        \hspace{0.2cm} synthetic dataset size & 20,000 \\
        \hspace{0.2cm} effective size with augmentation & 500,000 \\
        \hline
        max signed ``jiggle'' noise added to  & \\
        \hspace{0.2cm} all params of new predator CNN & $\pm$0.003 \\
        \hline
        static quality metric: trials per prey & 10\\
        \hline
    \end{tabular}
    \caption{Key simulation parameters. Details in source code.}
    \label{table:key_simulation_parameters} 
\end{table}
\end{footnotesize}


\subsection{Background Image Sets}
\begin{minipage}{\linewidth}
Names and descriptions for the sets of background images used in this paper, see Section \nameref{subsec:background_sets}. Each set is composed of several photographs of a similar natural scene.
\par
\begin{table}[H]
    \footnotesize
    \centering
    \begin{tabular}{ |l|l|c|c| }
        \hline
        \textbf{name} & \textbf{description} & \textbf{photos} & \textbf{figures} \\ 
        \hline
        backyard\_oak &
            under canopy of California live oak (\textit{Quercus agrifolia}) &
            12 & \ref{fig:backyard_oak_4x} \\
        \hline
        bean\_soup\_mix &
            mixture of dried beans from grocery store &
            4 & \ref{fig:bean_soup_mix_4x} \\
        \hline
        jans\_oak\_leaves &
            white oak leaf litter (Jan Allbeck, Fairfax, Virginia) &
            6 & \ref{fig:jans_oak_leaves_4x} \\
        \hline
        kitchen\_granite &
            polished granite counter-top in our kitchen &
            6 & \ref{fig:kitchen_granite_4x} \\
        \hline
        mbta\_flowers &
            flowers (impatiens?) near MBTA Northeastern stop &
            4 & \ref{fig:predator_responses}, \ref{fig:mbta_flowers_4x} \\
        \hline
        michaels\_gravel &
            gravel bed in neighbor’s front yard &
            4 & \ref{fig:teaser}, \ref{fig:predator_cnn}, \ref{fig:michaels_gravel_4x} \\
        \hline
        oak\_leaf\_litter &
            fallen oak leaves on edge of road &
            6 & \ref{fig:time_sequence} \\
        \hline
        oxalis\_sprouts &
            sprouts of Oxalis push through leaf litter after first rain &
            5 & \ref{fig:teaser} \\
        \hline
        plum\_leaf\_litter &
            fallen leaves from plum and other trees, near sunset &
            5 & \ref{fig:teaser}, \ref{fig:plum_leaf_litter_4x} \\
        \hline
        redwood\_leaf\_litter &
            dried redwood leaf litter collected in a roadside gutter &
            4 & \ref{fig:redwood_leaf_litter_4x} \\
        \hline
        rock\_wall &
            ``dry stack'' retaining wall in a neighbor's front yard &
            14 & \ref{fig:rock_wall_4x} \\
        \hline
        tiger\_eye\_beans &
            dried heirloom ``tiger eye'' beans from farmers market  &
            5 & \ref{fig:simulation_overview} \\
        \hline
        tree\_leaf\_blossom\_sky &
            small trees (branch, leaf, and blossom) sky background &
            5 & \ref{fig:teaser}, \ref{fig:tree_leaf_blossom_sky_4x} \\
        \hline
        yellow\_flower\_on\_green &
            ``Scot's broom'' (or ``French broom''?) in neighbor's yard &
            6 & \ref{fig:yellow_flower_4x} \\
        \hline
    \end{tabular}
    \label{table:background_sets}
\end{table}
\end{minipage}


\subsection{Details of Pre-Trained Predator Model}
\label{sec:pretrained_predator_details}
\vspace{0.2cm}
\begin{minipage}{\linewidth}
The pre-trained predator visual system, shown in Fig. \ref{fig:predator_cnn}, is a Keras TensorFlow CNN model with about 3.2 million parameters. Its input is a 128$\times$128 pixel RGB image (128$\times$128$\times$3 scalar values) and its output is an \textit{xy} location where the model estimates the most conspicuous prey is centered. Here is its Keras \texttt{model.summary()}:
\par

\vspace{0.2cm}

\begin{minipage}{\linewidth-1.1cm}
\hspace*{1cm}
\begin{minipage}{\linewidth-1.1cm}
\begin{small}
\begin{verbatim}

Model: "sequential"
_________________________________________________________________
 Layer (type)                Output Shape              Param #
=================================================================
 conv2d (Conv2D)             (None, 128, 128, 16)      1216
 dropout (Dropout)           (None, 128, 128, 16)      0
 conv2d_1 (Conv2D)           (None, 64, 64, 32)        12832
 dropout_1 (Dropout)         (None, 64, 64, 32)        0
 conv2d_2 (Conv2D)           (None, 32, 32, 64)        51264
 dropout_2 (Dropout)         (None, 32, 32, 64)        0
 conv2d_3 (Conv2D)           (None, 16, 16, 128)       204928
 dropout_3 (Dropout)         (None, 16, 16, 128)       0
 conv2d_4 (Conv2D)           (None, 8, 8, 256)         819456
 dropout_4 (Dropout)         (None, 8, 8, 256)         0
 flatten (Flatten)           (None, 16384)             0
 dense (Dense)               (None, 128)               2097280
 dropout_5 (Dropout)         (None, 128)               0
 dense_1 (Dense)             (None, 32)                4128
 dense_2 (Dense)             (None, 8)                 264
 dense_3 (Dense)             (None, 2)                 18
=================================================================
Total params: 3,191,386
Trainable params: 3,191,386
Non-trainable params: 0
\end{verbatim}
\end{small}
\end{minipage}
\end{minipage}
\end{minipage}
\par


\vspace{0.5cm}

\subsection{\texsyn{} \texttt{c++} Code for Figure \ref{fig:TexSyn_overview}}
\label{sec:cpp_code}
\begin{minipage}{\linewidth}
\begin{small}
\begin{verbatim}

    Uniform white(1);
    Uniform gray(0.1);
    Uniform blue(0, 0, 1);
    Uniform green(0, 1, 0);
    LotsOfSpots spots(0.9, 0.05, 0.3, 0.02, 0.02, blue, white);
    Grating stripes(Vec2(), green, Vec2(0.1, 0.2), gray, 0.3, 0.5);
    NoiseWarp warp_stripes(1, 0.1, 0.7, stripes);
    LotsOfSpots spots2(0.9, 0.05, 0.3, 0.02, 0.02, stripes, white);
    Grating stripes2(Vec2(), green, Vec2(0.1, 0.2), spots, 0.3, 0.5);
    NoiseWarp warp_all(1, 0.1, 0.7, stripes2);

\end{verbatim}
\end{small}
\end{minipage}


\vspace{0.5cm}
\subsection{Additional command line arguments to \texsyn{} for simulation runs}
\label{sec:texsyn_cmd_line_args}
\begin{minipage}{\linewidth}
\begin{small}
\begin{verbatim}

    background image directory (required)
    output directory (defaults to .)
    background scale (defaults to 0.5)
    random seed (else: default seed)
    window width (defaults to 1200)
    window height (defaults to 800)
    individuals (defaults to 120)
    subpopulations (defaults to 6)
    max init tree size (defaults to 100)
    min crossover tree size (default max init tree size  * 0.5)
    max crossover tree size (default max init tree size  * 1.5)
\end{verbatim}
\end{small}
\end{minipage}

\newpage

\begin{figure}[t]
    \igfour{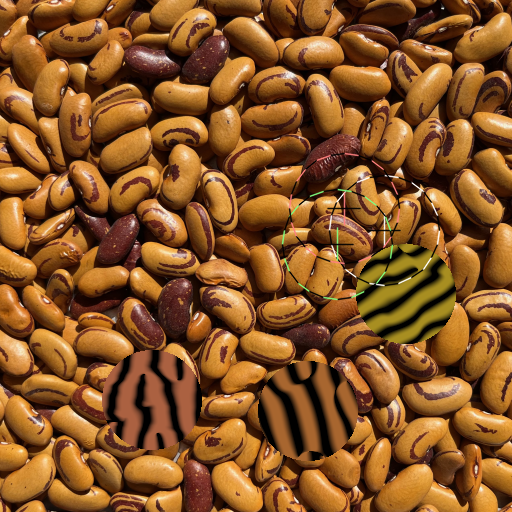}
    \hfill
    \igfour{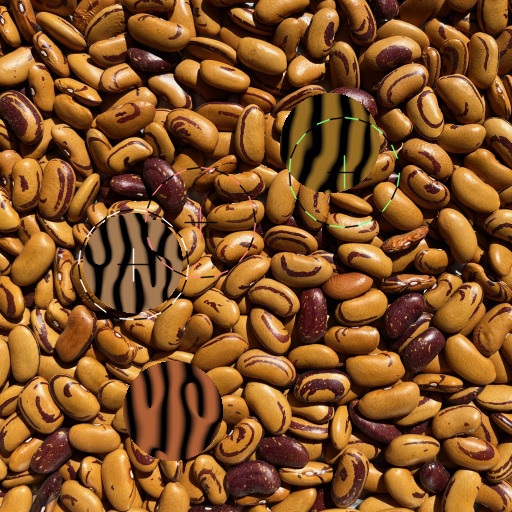}
    \hfill
    \igfour{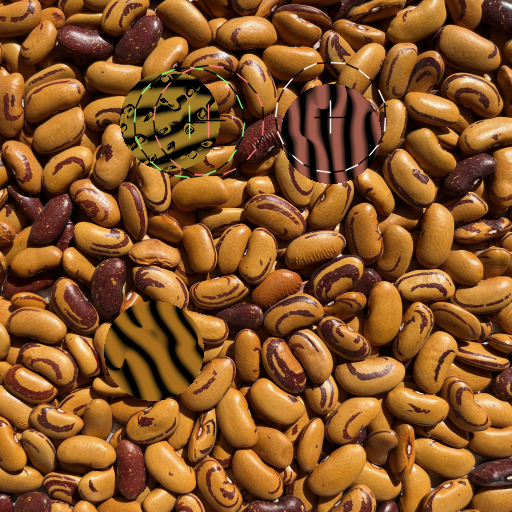}
    \hfill
    \igfour{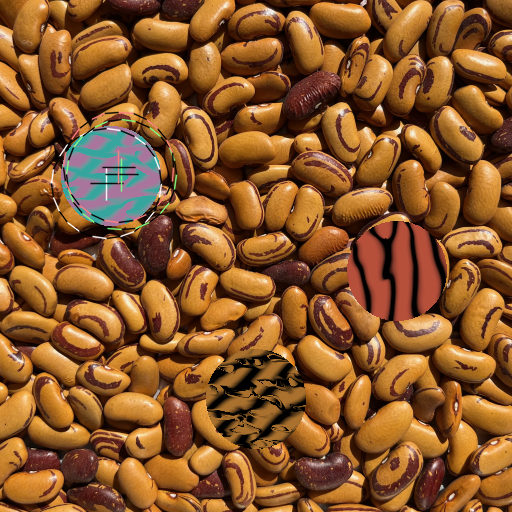}
    \caption{Four tournament images, plus predator responses as crosshairs, from run {\runID tiger\_eye\_beans\_20220903\_1401}.}
    \label{fig:tiger_eye_beans_4x}
\end{figure}


\subsection{Additional Samples of Predator Responses}
\label{sec:additional_predator_responses}
Similar to Fig. \ref{fig:predator_responses}, the examples in Figure \ref{fig:tiger_eye_beans_4x} show the prediction output of all three predators in each tournament. The crosshairs are ordered by least ``aim error'' with the best drawn in black and white, second in black and green, and worst in black and red. From left to right: (a) all three predators miss all three prey, (b) best is near center of one prey, second is off center but still inside another prey, and third fails to find any prey, (c) all three succeed, best inside one prey, the other two inside another prey, and (d) all three predators select the same prey which has especially conspicuous coloration.
\par


\end{document}